\documentclass[11pt,a4paper,oneside]{article}
\pdfoutput=1
\usepackage{jheppub}
\usepackage{hyperref}
\usepackage{graphicx,xcolor}
\usepackage{sectsty,amssymb,amsmath}
\usepackage{booktabs}
\usepackage[utf8]{inputenc} 
\usepackage{subcaption}
\usepackage{siunitx}
\usepackage{cleveref}
\usepackage{placeins}

\newcommand{\cO}{\mathcal{O}}
\newcommand{\cL}{\mathcal{L}}

\newcommand{\upd}{\mathrm d}                                          % Upright d for integrals
\newcommand{\ba}[1]{\bigg(#1\bigg)}                                   % ()
\newcommand{\bb}[1]{\bigg[#1\bigg]}                                   % []
\DeclareMathOperator\arsinh{arsinh}

\newcommand{\abs}[1]{\left\lvert#1\right\rvert}

\newcommand{\vev}[1]{{\langle #1 \rangle}}

\begin{document}

%=============================================================================
\title{Filtered Baryogenesis}

\author[a]{Michael~J.~Baker,}    \emailAdd{michael.baker@unimelb.edu.au}
\author[b]{Moritz~Breitbach,}    \emailAdd{breitbach@uni-mainz.de}
\author[c,b]{Joachim~Kopp,}      \emailAdd{jkopp@cern.ch}
\author[b]{Lukas~Mittnacht,}     \emailAdd{lmittna@uni-mainz.de}
\author[d]{Yotam~Soreq}          \emailAdd{soreqy@physics.technion.ac.il}

\affiliation[a]{ARC Centre of Excellence for Dark Matter Particle Physics, School of Physics, The University of Melbourne, Victoria 3010, Australia}
\affiliation[b]{PRISMA Cluster of Excellence and
                Mainz Institute for Theoretical Physics,
                Johannes Gutenberg-Universit\"{a}t Mainz, Germany}
\affiliation[c]{Theoretical Physics Department, CERN, Geneva, Switzerland}
\affiliation[d]{Physics Department, Technion—Israel Institute of Technology, Haifa 3200003, Israel}

\date{\today}

\preprint{CERN-TH-2021-219}
%=============================================================================

\abstract{
  We propose a new mechanism to simultaneously explain the observed dark matter abundance and the baryon asymmetry of the Universe. 
  The mechanism is based on the Filtered Dark Matter scenario, where dark matter particles acquire a large mass during a first-order phase transition.  
  This implies that only a small fraction of them are energetic enough to enter the advancing true vacuum bubbles and survive until today, while the rest are reflected and annihilate away quickly. 
  We supplement this scenario with a CP-violating interaction, which creates a chiral asymmetry in the population of dark matter particles. 
  In the false vacuum phase, a portal interaction quickly converts the dark sector chiral asymmetry into a Standard Model lepton asymmetry.
  The lepton asymmetry is then partially converted to a baryon asymmetry by standard electroweak sphaleron processes. 
  We discuss the dependence of the generated asymmetry on the parameters of the model for two different portal interactions and demonstrate successful baryogenesis for both.
  For one of the portals, it is also possible to simultaneously explain the observed dark matter abundance, over many orders of magnitude in the dark matter mass.
}

%-----------------------------------------------------------------------------
\maketitle
%-----------------------------------------------------------------------------

%-----------------------------------------------------------------------------
\section{Introduction}
\label{sec:intro}
%-----------------------------------------------------------------------------

Despite the manifold successes of the Standard Model~(SM) of particle physics, it does not provide an explanation for the dark matter seen in the Universe, nor does it account for the Baryon Asymmetry of the Universe~(BAU).  
In this work we explore a mechanism which sets the dark matter abundance and the baryon asymmetry at the same time, providing a joint explanation to these disparate puzzles. 

To account for dark matter, we utilise the recently proposed Filtered Dark Matter scenario~\cite{Baker:2019ndr, Chway:2019kft}, which is an alternative to the canonical Weakly Interacting Massive Particle~(WIMP) paradigm for dark matter~\cite{Battaglieri:2017aum}.
The idea behind Filtered Dark Matter is that the dark matter abundance is set during a first-order cosmological phase transition.
As is well known, such a phase transition proceeds via the formation and subsequent expansion of true-vacuum bubbles in a Universe that is otherwise still in a false vacuum.
In the Filtered Dark Matter scenario, dark matter particles acquire a large mass during the phase transition. 
This means that only particles in the Boltzmann tails of the dark matter velocity distribution have sufficient energy to enter the bubbles, while most of the particles reflect off the advancing bubble wall, remain massless, and annihilate.
In contrast, the few dark matter particles that manage to enter the bubble freeze-out instantaneously due to their large masses and can no longer annihilate.
Overall, only a small fraction of the dark matter particles survive until today, in agreement with observations.

In the present article, we show that the Filtered Dark Matter mechanism can be extended to accommodate baryogenesis.
Specifically, we consider a phase transition where a new scalar $\phi$ obtains a vacuum expectation value above the electroweak scale, and we argue that an additional CP-violating interaction between the dark matter particle $\chi$ and $\phi$ can make the dark matter--bubble wall interactions chirally asymmetric: $\chi$ particles of one chirality will have a slightly higher reflection probability than those of the opposite chirality.
As a result, a chiral asymmetry is generated close to the bubble wall, reminiscent of the SM chiral asymmetry generated in electroweak baryogenesis models \cite{%
    Kuzmin:1985mm,        % first paper on EWBG
    Shaposhnikov:1986jp,  %
    Shaposhnikov:1987tw,  %
    Carena:1996wj,
    Huber:2006wf,         % EWBG in the MSSM
    Vaskonen:2016yiu,     % GW from EWBG phase transition
    Garbrecht:2018mrp,    % calculational methods
    deVries:2018tgs,      % the role of leptons in EWBG
    Cline:2020jre,        % high wall velocities
    Laurent:2020gpg,      % fluid equations in the case of fast bubble walls
    Fuchs:2020pun}        % analytic techniques for EWBG
(see refs.~\cite{Cohen:1993nk, Cline:2006ts, Morrissey:2012db} for review articles on this topic).
However, while SM sphaleron processes convert the chiral asymmetry into a particle--antiparticle asymmetry in electroweak baryogenesis, we will use a lepton-number-violating portal interaction to do so.
As the $\phi$ bubble wall sweeps past, the large dark matter mass ensures that this portal does not wash out the newly-generated lepton asymmetry.
Electroweak sphalerons finally convert part of the lepton asymmetry into a baryon asymmetry.

The filtering effect proposed in refs.~\cite{Baker:2019ndr, Chway:2019kft} has been applied to baryogenesis in ref.~\cite{Arakawa:2021wgz} where reflected particles build up in the false vacuum phase, leading to an enhanced asymmetry due to annihilations (this work does not, however, simultaneously produce dark matter).
Baryogenesis due to the production of heavy particles from relativistic bubble walls has been demonstrated in ref.~\cite{Azatov:2021irb}.
The idea of generating the observed baryon asymmetry by first generating an asymmetry in a dark sector has previously been discussed in various contexts.
The models proposed in refs.~\cite{Dutta:2006pt, Shelton:2010ta, Hall:2019ank} move the concept of electroweak baryogenesis (including the sphalerons) entirely to the dark sector by introducing a non-Abelian gauge symmetry there.
Ref.~\cite{Cline:2017qpe} adds a scalar singlet to the SM Higgs sector, thus rendering the electroweak phase transition first order~\cite{Choi:1993cv, Espinosa:2011ax, Cline:2012hg}.  
In this model, a chiral asymmetry is first generated in the dark sector and then transferred to the SM via a coupling between the dark matter, the $\tau$ lepton, and a new, inert, Higgs doublet. 
Apart from the very different portal interaction, this model relies on conventional thermal freeze-out to set the dark matter relic density, while in our case the phase transition is responsible for both baryogenesis and the dark matter abundance.
Other works relating a dark sector asymmetry to the baryon asymmetry in the SM sector include refs.~\cite{%
    Nussinov:1985xr,     % first suggestion of asymmetric DM (with technibaryons)
    Barr:1990ca,         % another early paper proposing asymmetric DM
    Barr:1991qn,         % if a DM quantum number is conserved in addition to B-L, SM sphalerons may not erase B+L
    Kaplan:2009ag,       % the asymmetric DM paper that set off the ADM craze in 2009-2011
    An:2009vq,           % a mirror sector and leptogenesis
    Haba:2010bm,         % DM asymmetry generated in particle decays, then transfer to the visible sector
    Davoudiasl:2010am,   % Hylogenesis (equal + opposite asymmetries in the dark and visible sectors)
    Buckley:2010ui,      % Xogenesis
    Gu:2010ft,           % B or L-conserving models where the missing B/L number is carried by DM
    Blennow:2010qp,      % Aidnogenesis: new non-Abelian symmetry common to DM and SM sectors
    Allahverdi:2010rh,   % moduli decay
    Dutta:2010va,        % ewk-like baryogenesis in dark sector, inspired by Dutta:2006pt
    Falkowski:2011xh,    % heavy RH neutrinos decay with CPV to both the visible and dark sectors
    Graesser:2011wi,     % various asymmetric DM scenarios
    Buckley:2011kk,      % effective operators for transferring the asymmetry
    Bell:2011tn,         % conserved "generalized B"; Affleck-Dine scalar dynamically breaks it
    Cheung:2011if,       % Affleck-Dine cogenesis
    MarchRussell:2011fi, % spontaneous baryogenesis in a dark sector, then transfer to the visible sector
    Cui:2011qe,          % asymmetry transfer via mass mixing that is switched off at some point
    Cui:2011ab,          % WIMPy baryogenesis miracle; asymmetry particle hidden
    Davoudiasl:2012uw,   % Review
    Unwin:2012rp,        % on the transfer of asymmetries between DM and SM sectors
    Cui:2012jh,          % B-violating decays of metastable particle
    Barr:2013tea,        % large DS gauge group convert primordial B or L asymmetry to DM asymmetry
    Servant:2013uwa,     % Higgsogenesis -- transferring asymmetry between SM and DM sectors via the Higgs
    Feng:2013wn,         % asymmetric inflaton decay into the DM sector, then transfer of asymmetry to SM sector
    Hall:2019ank,        % Baryogenesis From a Dark First-Order Phase Transition
    Hall:2019rld}.       % Asymmetric Matters from a Dark First-Order Phase Transition

This paper is structured as follows:~in \cref{sec:setup}, we discuss our extension of the Filtered Dark Matter model in more detail and set down our notation.
In \cref{sec:transeqs} we introduce the transport equations that determine the dark sector chiral asymmetry and the SM baryon asymmetry.
We also explain how we solve these equations numerically.  
We present our main results in \cref{sec:results}, where we highlight the model's behaviour at two benchmark parameter points before showing the results of parameter scans.  These scans illustrate that the mechanism is able to explain both the dark matter abundance and the baryon asymmetry in the Universe over wide ranges of parameter space.
We summarise and conclude in \cref{sec:conclusions}.
A possible connection to neutrino masses is given in \cref{sec:app-inverse-seesaw} while technical details are discussed in appendices~\ref{sec:app-extended-transport} to~\ref{sec:app-diff-approx}.

%-----------------------------------------------------------------------------
\section{Model and General Mechanism}
\label{sec:setup}
%-----------------------------------------------------------------------------

Our starting point is the Filtered Dark Matter scenario from ref.~\cite{Baker:2019ndr}, which we briefly review here.
Filtered Dark Matter is based on the assumption that the Universe undergoes a first-order phase transition from a false-vacuum state in which the dark matter particle $\chi$ is massless or very light, to a true-vacuum state in which the mass of $\chi$, $m_\chi$, is large compared to the phase transition temperature.
Conservation of energy then implies that only $\chi$ particles with kinetic energy $E_\chi \gtrsim m_\chi$ can pass through the advancing bubble walls.
The majority of $\chi$ particles are reflected, and quickly annihilate.
The particles which pass through the wall obtain a large mass which, combined with their low number density, means that their interaction rate is no longer large enough to keep them in thermal equilibrium.
These $\chi$ particles are thus frozen out and survive until the present day.  
The resulting relic abundance is approximately given by~\cite{Baker:2019ndr}
\begin{align}
  \Omega_\mathrm{DM} h^2
    &\simeq 
    0.17 \, 
    \bigg( \frac{T_n}{\text{TeV}} \bigg)
    \bigg( \frac{m_\chi / T_n}{30} \bigg)^{\!5/2}
    \frac{e^{- m_\chi / T_n}}{e^{-30}} \, ,
  \label{eq:Omega_chi}
\end{align}
where $T_n$ is the phase transition temperature. 
For definiteness, we take dark matter to be a gauge-singlet Dirac fermion $\chi$, charged under a $\mathbb{Z}_2$ symmetry.

We assume that the phase transition is driven by a real gauge-singlet scalar field $\phi$, whose vacuum expectation value~(vev), $v_\phi$, vanishes in the false vacuum, but is non-zero in the true vacuum.
The mass $m_\chi$ of the dark matter particles is induced by $v_\phi$ through a Yukawa coupling,
\begin{align}
    \cL \supset&\,
	{-}y_\chi \phi \overline{\chi}\chi 
	\quad \to \quad
	{-}y_\chi v_\phi \overline{\chi}\chi 
	\sim {-}m_\chi \overline{\chi}\chi\,.
\end{align}
As we will discuss below, $m_\chi$ receives additional contributions from higher-dimensional operators, in particular from the CP-violating term which we will add.
We assume the bare mass term of $\chi$ to be negligible.
Here and in the following, we use $m_\chi$ and $v_\phi$ to denote the mass of $\chi$ and the vev of $\phi$, respectively, deep inside the true vacuum.
When discussing the transition between the two vacua at the bubble wall, we will write $m_\chi(z)$ and $v_\phi(z)$ for the $z$-dependent quantities.

For this mechanism to set the correct dark matter relic abundance, only very few particles should be able to enter the bubble. Similar to ref.~\cite{Baker:2019ndr}, this leads to the requirement $m_\chi \gg T_n$, which in turn requires a large order parameter $v_\phi \gg T_n / y_\chi$.
While we remain agnostic about the precise form of the scalar potential, $V(H,\phi)$, we note that this feature is realised for instance in quasi-conformal or dilaton-like setups~\cite{Hambye:2018qjv,DelleRose:2019pgi}.

The scalar potential contains the Higgs portal coupling,
\begin{align}
    V(H,\phi) \supset \beta \, \phi^2 |H|^2 \,,
    \label{eq:V}
\end{align}
through which the dynamics of electroweak symmetry breaking can be modified.
This is not necessarily a problem for the success of Filtered Baryogenesis, but would greatly complicate calculations.
Therefore, we will assume that electroweak symmetry breaking remains unchanged.
This may require fine-tuning between the Higgs portal and the tree-level mass term of the SM Higgs field, which is a reflection of the well-known hierarchy problem inherent to scalar potentials with widely separated energy scales.

The two new ingredients added in this work are a CP-violating coupling between $\chi$ and $\phi$ (which leads to CP-violating interactions between $\chi$ particles and the bubble walls), and a lepton-number-violating portal interaction between $\chi$ and SM particles.
The portal interaction will convert a chiral asymmetry in $\chi$ to a particle--antiparticle asymmetry in the SM sector.
(Note that the Higgs portal coupling, \cref{eq:V}, cannot transmit chiral asymmetries between the sectors.)

Overall, the dark sector Lagrangian takes the form
\begin{align}
   \cL = \cL_\chi + \cL_\text{p} + \cL_5 - V(H,\phi) \,.
   \label{eq:L}
\end{align}
We now discuss $\cL_\chi$, $\cL_\text{p}$, and $\cL_5$ in turn.
The $\chi$--$\phi$ couplings that contain the CP-violating term are
\begin{align}
    \cL_{\chi} = -y_\chi \ba{
                     \phi + \frac{T^\chi_R + i \,T^\chi_I}{v_\phi} \phi^2
                 } \overline{\chi_R} \chi_L + h.c. \,,
	\label{eq:LY}
\end{align}
with real dimensionless parameters $y_\chi$, $T^\chi_R$, and $T^\chi_I$.
Without loss of generality, we use the vev of $\phi$ deep inside the true vacuum phase as the suppression scale of the dimension-5 coupling.
For simplicity, we take the CP-violating phase in the dimension-5 operator to be maximal, that is, we choose $T^\chi_R = 0$. 
Since $m_\chi \sim y_\chi v_\phi$ and since we will be interested in Yukawa couplings of order one, using an EFT approach for this operators is valid as long as $|T^\chi_I| \lesssim 1$.

For the portal interaction, $\cL_{\rm p}$, we require a lepton-number violating operator to convert a chiral asymmetry in $\chi$ into a lepton asymmetry.  There are various options, and we consider two possibilities. 
Perhaps the simplest is
\begin{align}
    \cL_{\text{p}} = \frac{1}{\Lambda_{\rm p}^2} \sum_{j=1,2,3} (\overline{N_R^j} N_R^{jc}) \,
                                                                (\overline{\chi_R}{\chi_L})
                   + \sum_{j,k=1,2,3} y_\nu^{jk} \overline{L^j} \tilde{H} N_R^k
                   + h.c. \,,
    \label{eq:portal-NR}
\end{align}
where $N_R^j$ are three generations of heavy right-handed Majorana neutrino fields carrying one unit of lepton number (which could be part of a seesaw scenario that explains the smallness of the SM neutrino masses), $\Lambda_{\rm p}$ is the suppression scale of the dimension-6 operator, which we assume to be the same for all generations, and the superscript $c$ denotes the charge-conjugation operation, $N_R^{jc} \equiv i \gamma^2 (N_R^j)^*$~\cite{Akhmedov:1999uz}.
In the following, we usually suppress flavour indices to make the notation more concise.
The $N_R$ portal could be generated, for instance, when a heavy scalar that couples to both $\chi$ and $N_R$ is integrated out.
We will refer to \cref{eq:portal-NR} as the ``dimension-6 $N_R$ portal''.
If the chirality-dependent permeability of the bubble walls creates, say, an overabundance of right-handed $\chi_\text{RH}$ and $\bar{\chi}_\text{RH}$ over left-handed $\chi_\text{LH}$ and $\bar{\chi}_\text{LH}$ in front of the wall,\footnote{For clarity we differentiate the chiral $\chi$ field operators, e.g.~$\chi_L$, from the particles of definite chirality they create and annihilate. The field operator $\chi_L$ annihilates $\chi_\text{LH}$ and creates $\bar\chi_\text{RH}$, while the Hermitian conjugate field operator $\overline{\chi_L}$ creates $\chi_\text{LH}$ and annihilates $\bar\chi_\text{RH}$.}
the dimension-6 $N_R$ portal will convert this asymmetry into a deficit of $N_\text{RH}$ compared to its left-handed antiparticle, $\bar{N}_\text{LH}$.
Through the Yukawa couplings $y_\nu \overline{L} \tilde{H} N_R$, each $N_\text{RH}$ and $\bar{N}_\text{LH}$ particle decays into a lepton and a Higgs boson, generating slightly fewer SM leptons than antileptons. 
In the following, we assume that $N_R^j$ couples only to the $j$-th generation of SM leptons and that the coupling strength is the same for all generations.
In other words, we assume that $y_\nu$ is proportional to the unit matrix.  
(Note that we work here in unbroken $SU(2)_L$, assuming that the phase transition responsible for baryogenesis occurs above the electroweak scale. Thus, $\tilde{H} = i \sigma^2 H^*$ contains four physical real scalar bosons. The lepton doublet $L$ contains, as usual, the left-handed neutrino and charged lepton fields.)
The lepton asymmetry is then partially converted into a baryon asymmetry by baryon-plus-lepton number ($B+L$) violating (but $B-L$ conserving) SM sphaleron processes.
For this mechanism to work, it is crucial that the lepton asymmetry carried intermittently by $N_R$ is not washed out.
This implies that other lepton number violating interactions -- in particular a possible Majorana mass term for $N_R$ -- must be suppressed until the time when $N_R$ decays, or when electroweak sphalerons freeze out, whichever happens earlier.
Satisfying this condition, while simultaneously explaining the smallness of neutrino masses, appears possible in inverse seesaw scenarios, see \cref{sec:app-inverse-seesaw}~\cite{Mohapatra:1986ks,CentellesChulia:2020dfh}.

A simple alternative to the dimension-6 $N_R$ portal is a dimension-8 operator connecting dark matter directly to the SM leptons: 
\begin{align}
	\cL_{\text{p}} = \frac{1}{\Lambda_{\text{p}}^4} \sum_{j=1,2,3} (\overline{L^{jc}} \tilde H^*) (\tilde H^\dagger L^j) (\overline{\chi_R} \chi_L)
	                 + h.c.
    \label{eq:portal-W}
\end{align}
As this is just a coupling of a dark matter current to the Weinberg operator, we will refer to this portal as the ``dimension-8 Weinberg portal''.
In the following, we will again suppress the generation index $j$ on the lepton doublets $L^j$.
The dimension-8 Weinberg portal directly converts the chiral asymmetry in the dark sector into a SM lepton--antilepton asymmetry, without requiring right-handed neutrinos (though the dimension-8 Weinberg portal can be generated from the dimension-6 $N_R$ portal upon integrating out the $N_R$ fields).

While a consistent effective field theory~(EFT) contains all possible operators up to a given dimension, we assume the leading contribution to the asymmetry transfer is given by the operator in \cref{eq:portal-NR} or in \cref{eq:portal-W}.  
Since operators not contributing to the transfer of the asymmetry, e.g., the usual Weinberg operator at dimension-5, do not appear in the relevant transport equations, we omit them from our analysis.
We do however include the dimension-5 operator
\begin{align}
    \cL_5 = \frac{\lambda_5}{\Lambda_{\text{p}}}
            (H^\dagger H)(\overline{\chi_R}\chi_L) + h.c. \,,
    \label{eq:L5}
\end{align}
which contributes to $\chi\bar\chi$ annihilation and can wash out the chiral asymmetry in $\chi$.
We will assume $\lambda_5$ to be real (that is, we do not consider the possibility that the washout rate is different for different chiralities).

\begin{figure}[t]
	\centering
	\includegraphics{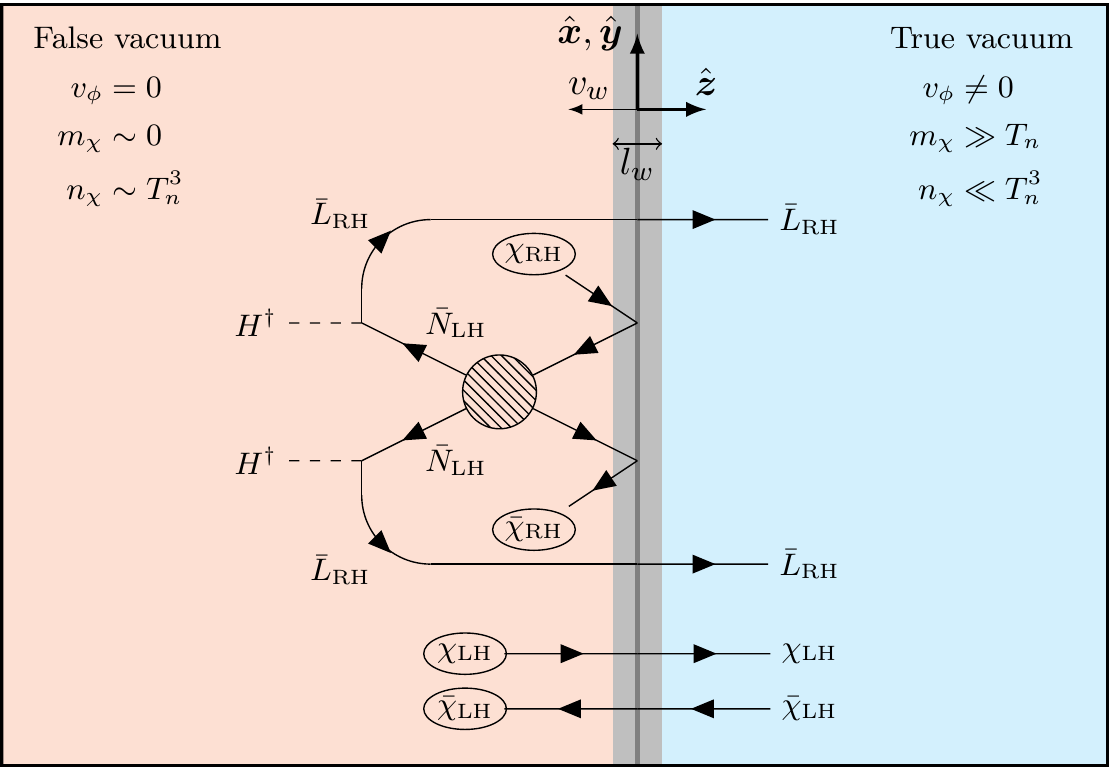}	
	\caption{Illustration depicting the generation of a SM lepton--antilepton asymmetry.  The particles $\chi_\text{RH}$ and $\bar\chi_\text{RH}$ are preferentially reflected from the wall and may be converted via the dimension-6 $N_R$ portal, \cref{eq:portal-NR}, into left-handed anti-neutrinos, $\bar N_\text{LH}$ (but not right-handed neutrinos, $N_\text{RH}$), which in turn decay to right-handed SM anti-leptons.  These anti-leptons are then converted into left-handed quarks by SM sphalerons. The particles $\chi_\text{LH}$ and $\bar\chi_\text{LH}$ are preferentially transmitted through the wall.  Ellipses around the $\chi$ particles indicate that they should be considered the initial states in the process.  See text for further details.
	}
	\label{fig:illustration}
\end{figure}

The general mechanism is summarised in \cref{fig:illustration}.
As the bubble wall, depicted as a vertical grey band, passes through space from right to left, it will encounter left-handed and right-handed $\chi$ particles and antiparticles.
While $\chi_\text{LH}$ and $\chi_\text{RH}$ are both predominantly reflected, due to the large gain in mass, the CP-violating coupling in \cref{eq:LY} means there is a slight preference for, say, $\chi_\text{RH}$ particles to be reflected and for $\chi_\text{LH}$ particles to be transmitted.
Similarly, there is a preference for $\bar\chi_\text{RH}$ antiparticles to be reflected and $\bar\chi_\text{LH}$ to be transmitted.
As such, opposite chiral asymmetries build up either side of the wall.
In front of the wall, the portal interaction converts the chiral asymmetry of $\chi$ into a particle--antiparticle asymmetry.
For the dimension-6 $N_R$ portal depicted in \cref{fig:illustration}, $\chi_\text{RH} \bar\chi_\text{RH} \to \bar N_\text{LH} \bar N_\text{LH}$ creates an $N_\text{RH}$--$\bar{N}_\text{LH}$ asymmetry, which is turned into a SM lepton--antilepton asymmetry through the Yukawa coupling between $N_R$, $L$ and $H$.
For the dimension-8 Weinberg portal, a SM lepton--antilepton asymmetry is directly generated by the portal interaction, without the detour via $N_R$.
Behind the wall, the large mass of $\chi$ washes out the chiral asymmetry of the transmitted $\chi$ particles.
However, since the portal interaction is suppressed in this region, it does not wash out the SM lepton asymmetry.
For both portals, electroweak sphalerons convert part of the resulting lepton asymmetry into a baryon asymmetry, which survives to the present day.

In the rest frame of the bubble wall, the system quickly reaches a steady state with the rate of generation of the chiral asymmetry in $\chi$ being matched by its conversion into a lepton asymmetry by the portal interaction and the conversion of the lepton asymmetry into a baryon asymmetry by electroweak sphalerons.
It is this steady-state solution that we will compute in \cref{sec:transeqs}.

Before ending this section, let us outline several conditions that must be met for the mechanism to be successful:  
\begin{enumerate}
    \item The CP-violating phase in \cref{eq:LY} must be large enough to ensure that a sufficient chiral asymmetry is generated.
    \item The phase transition must occur before the electroweak phase transition, so electroweak sphalerons can efficiently convert a lepton asymmetry into a baryon asymmetry.
\end{enumerate}

In addition, generating the observed dark matter abundance via the Filtered Dark Matter mechanism requires that:
\begin{enumerate}
    \item $\chi$ must be in equilibrium in the false vacuum ($\Gamma^\text{false}_{\chi\bar\chi\leftrightarrow \phi\phi}>H$) to ensure efficient annihilation of reflected $\chi$ particles.\footnote{Strictly speaking, the time it takes $\chi$ particles to annihilate should be smaller than the time it takes the phase transition to complete, which is highly model-dependent and may be much less than one Hubble time. Otherwise, an overdensity of $\chi$ particles will accumulate in the false vacuum. This overdensity will slow down the advancing bubble walls, delaying the completion of the phase transition. Moreover, it will lead to more $\chi$ particles entering the true vacuum, modifying the relation between the model parameters and the dark matter relic abundance, \cref{eq:Omega_chi}. See refs.~\cite{Baker:2021nyl,Baker:2021sno} for more details on the dynamics and implications of a $\chi$ overdensity.}
    
    \item $\chi$ must drop out of equilibrium in the true vacuum ($\Gamma^\text{true}_{\chi\bar\chi\leftrightarrow \phi\phi}<H$).
    If this were not the case, the dark matter abundance would be set by conventional thermal freeze-out at a later time and would not be determined by the phase transition.
    
    \item The relevant couplings must satisfy perturbative unitarity constraints ($y_\chi$, $\beta < \sqrt{4\pi}$).
    
    \item For the approximations employed in ref.~\cite{Baker:2019ndr} to be valid, $\phi$ must remain in equilibrium throughout the phase transition.
    This means that its interaction rate with the SM bath must be faster than the Hubble rate  ($\Gamma_{\phi\phi\leftrightarrow H^\dagger H} > H$) and that it must annihilate faster than it is produced by $\chi$ ($\Gamma_{\phi\phi\leftrightarrow H^\dagger H} > \Gamma_{\chi\bar\chi\leftrightarrow \phi\phi}$).
\end{enumerate}

Since the coupling that keeps $\phi$ in equilibrium is the Higgs portal, $\beta\,\phi^2 |H|^2$, the latter condition places a lower limit on $\beta$.
Interestingly, this lower limit on $\beta$ leads to a lower limit on the cross-section for dark matter--nucleon scattering via scalar exchange, which is probed in direct detection experiments.
The spin-independent dark matter--nucleon cross section, mediated by $t$-channel scalar exchange and taking into account $\phi$--$H$ mixing, is given by~\cite{Matsumoto:2018acr,Escudero:2016gzx}
\begin{align}
    \sigma_{\chi N}^{\rm SI} =&\,
    \frac{\beta^2f_q^2m_N^4m_\chi^4}{4\pi\,m_h^4m_\phi^4 (m_N+m_\chi)^2} \,,
    \label{eq:sigma-dd}
\end{align}
where $f_q$ parameterises the Higgs--nucleon coupling, $m_N$ is the nucleon mass and $m_h$ is the mass of the SM Higgs boson.  
For our numerical results in \cref{sec:results} we take $f_q = 0.326$~\cite{Djouadi:2011aa}.

%-----------------------------------------------------------------------------
\section{Solving the Transport Equations} 
\label{sec:transeqs}
%-----------------------------------------------------------------------------

To quantify the baryon asymmetry that can be produced in our scenario, we need to solve a coupled set of particle transport equations.  
In particular, we need to track the particle--antiparticle asymmetries in 
(i)~the dark matter particle $\chi$; 
(ii)~the SM lepton doublets that are produced by the portal interaction, either directly for the dimension-8 Weinberg portal, or via $N_R$ for the dimension-6 $N_R$ portal; 
(iii)~the SM quark doublets for which an asymmetry is generated by weak sphalerons; 
(iv)~the Higgs doublet which also participates in the portal interaction.

The particle--antiparticle asymmetries for the chiral components of $\chi$ are defined in terms of the corresponding number densities as $\chi \equiv n_{\chi_\text{LH}} - n_{\bar\chi_\text{RH}}$, $\bar\chi \equiv n_{\chi_\text{RH}} - n_{\bar\chi_\text{LH}}$. 
Note that $\chi = -\bar\chi$ at all times because our Lagrangian possesses a global $U(1)$ symmetry that protects $\chi$ number, and because we assume a $\chi$--$\bar\chi$-symmetric initial state.
The asymmetries $\ell$ and $q$ for the SM $SU(2)$ doublet leptons and quarks are defined analogously for each generation.  
Neglecting all SM Yukawa couplings and using the fact that the new interactions we introduced are flavor-universal, we can treat all three fermion generations equally. 
As the strong sphaleron rate is very fast compared to the electroweak sphaleron rate that generates the quark asymmetry, the asymmetries in the $SU(2)$ singlet quarks $u$, $d$, $c$, $s$, $t$, $b$ do not need to be tracked independently, but can just be set equal to $q/2$.  
Making use of this relation and taking into account all three generations, the net baryon number is $6q/3$.
The charged lepton singlets, on the other hand, do not acquire an asymmetry, so the total lepton asymmetry is $3\ell$. 
For the Higgs bosons in the unbroken electroweak-symmetry phase, the definition of the asymmetry is $h \equiv n_H - n_{H^\dag}$.  
Note that, to shorten the notation and since the usage should be clear from context, we denote the asymmetries by the same or similar symbols as the fields themselves.   
Note also that $\ell$ describes the asymmetry in a left-handed lepton doublet.  That is, the asymmetry between the left-handed charged leptons and right-handed charged antileptons in that generation, \emph{in addition to} the corresponding neutrino asymmetry.  
The situation is analogous for the asymmetries in the $SU(2)$ doublets, $q$ and $h$.

While transport equations can be derived from a Boltzmann equation, the transport equation for each species is effectively a diffusion equation supplemented with source and interaction terms that describe the processes generating the asymmetry and shuffling it around between different species. Although a diffusion equation usually describes the evolution of a system in space and time, we can drop the time coordinate if we go to the rest frame of the wall and use the fact that the system quickly reaches a stationary state in that frame.
Moreover we work with only a single spatial coordinate, $z$, since we assume asymmetries are generated over distance scales that are much smaller than the diameter of the bubble.
We therefore treat the bubble wall as locally planar, which implies translation invariance with respect to the $x$ and $y$ coordinates.

In the rest frame of the bubble wall, the transport equations for the model with the dimension-6 $N_R$ portal operator, \cref{eq:portal-NR}, are (see e.g. refs.~\cite{deVries:2017ncy, DeVries:2018aul, Fuchs:2020uoc, Fuchs:2020pun})
\begin{align}
    \begin{split}
        \partial_\mu j_\chi^\mu
        &= 	 3 \Gamma_{\rm p} \tilde\mu_{\rm p}
            +   (\Gamma_{\rm 5}+\tilde \Gamma_5) \tilde\mu_{\rm 5}
            +   \Gamma^\chi_M \tilde\mu^{\chi}_M
            + S_{\chi} \,,\\[1ex]
        \partial_\mu j_N^\mu
        &=   2 \Gamma_{\rm p} \tilde\mu_{\rm p} - \Gamma_Y^N\tilde\mu_Y^N\,,\\[1ex]
        \partial_\mu j_\ell^\mu
        &= \Gamma_Y^N\tilde\mu_Y^N - \Gamma_\text{ws} \tilde\mu_\text{ws}\,,\\[1ex]
        \partial_\mu j_h^\mu
        &= 3\Gamma_Y^N\tilde\mu_Y^N\,,\\[1ex]
        \partial_\mu j_q^\mu
        &= - \frac{3}{2}\Gamma_\text{ws} \tilde\mu_\text{ws} \,,
    \end{split}
    \label{eq:transport-eq-dim-6}
\end{align}
and the transport equations for the dimension-8 Weinberg portal operator, \cref{eq:portal-W}, are
\begin{align}
    \begin{split}
        \partial_\mu j_\chi^\mu
        &= 	 3 \Gamma_{\rm p} \tilde\mu_{\rm p}
            +   (\Gamma_{\rm 5}+\tilde \Gamma_5) \tilde\mu_{\rm 5}
            +   \Gamma^\chi_M \tilde\mu^{\chi}_M
            + S_{\chi} \,,\\[1ex]
        \partial_\mu j_\ell^\mu
        &=   2 \Gamma_{\rm p} \tilde\mu_{\rm p}
        - \Gamma_\text{ws} \tilde\mu_\text{ws}\,,\\[1ex]
        \partial_\mu j_h^\mu
        &= 6 \Gamma_{\rm p} \tilde\mu_{\rm p} \,,\\[1ex]
        \partial_\mu j_q^\mu
        &= - \frac{3}{2}\Gamma_\text{ws} \tilde\mu_\text{ws}\,.
    \end{split}
    \label{eq:transport-eq-dim-8}
\end{align}
Here, $j_a^\mu$ is the 4-current density of the asymmetry of a given species (with the asymmetry $a \equiv j_a^0$), the interaction terms of the form $\Gamma \tilde\mu$ (where $\Gamma$ is an interaction rate and $\tilde\mu$ are combinations of effective chemical potentials, see below) describe the processes that shuffle the particle--antiparticle asymmetry between species, and $S_\chi$ is the CP-violating source term.
We have checked that our numerical results do not change significantly if we treat the third-generation fermions separately from their first- and second-generation counterparts \cite{Fuchs:2020pun}, see \cref{sec:app-extended-transport}.  We now describe the different terms in \cref{eq:transport-eq-dim-6,eq:transport-eq-dim-8} one by one.

The divergence can be written as
\begin{align}
    \partial_\mu j_a^\mu &
	           = \frac{\partial a}{\partial t} + \vec\nabla \cdot \vec{j}_a
	           = [v_w - D_a'(z)] a'(z) - D_a(z) a''(z) - D_a''(z) a(z) \,,
	\label{eq:fokker-planck}
\end{align}
where $v_w$ is the bubble wall velocity and we have used a generalised Fick's law to write the asymmetry current as $\vec{j}_a = -\vec\nabla(D_a(z) a)$.  We have moreover used translational invariance in $x$ and $y$.
We explain in \cref{sec:app-diff-coeff} how the diffusion coefficient of species~$a$, $D_a(z)$, can be calculated.
We take $D_a$ to be constant for the SM particles, while $D_\chi$ depends on $z$ through $m_\chi(z)$ (which changes significantly as a $\chi$ particle moves through the bubble wall).  
Note that the terms $D_a'(z) a'(z)$ and $D_a''(z)$ do not usually appear in the literature on electroweak baryogenesis, where all $D_a$ are typically taken to be independent of $z$. 
Assuming a stationary solution, we have rewritten the time derivative of the asymmetry $a$ in \cref{eq:fokker-planck} as the bubble wall velocity, $v_w$, times the spatial derivative of $a$. 
This term then describes the bulk particle flux in the wall's rest frame.  
Note that, with these definitions, $\partial_\mu j_a^\mu = 0$ is just the Fokker--Planck equation in one spatial dimension for a stationary system.

The conversion rates between particle species are given by the $\Gamma$'s. 
On dimensional grounds we assume the rate for the $d$-dimensional portal interaction, which directly or indirectly gives rise to the process $\chi\bar\chi \to 2L + 2H$,  to be
\begin{align}
    \Gamma_{\rm p}=T\ba{\frac{T}{\Lambda_{\rm p}}}^{\!2(d-4)} {\exp}\ba{{-}\frac{m_\chi(z)}{T}} \,,
\end{align}
at temperature $T$ for each lepton generation.  
The exponential factor accounts for the Boltzmann suppression of the $\chi$ abundance inside the broken phase ($z>0$). 
This suppression is crucial for avoiding washout of the generated asymmetries after the phase transition is over. 
By similar dimensional arguments, we write the rate of washout due to the dimension-5 operator in \cref{eq:L5} as
\begin{align}
    \Gamma_5 
    &= T\ba{\lambda_5\frac{T}{\Lambda_{\rm p}}}^{\!2} 
    {\exp}\ba{{-}\frac{m_\chi(z)}{T}} \,,
\end{align}
which describes the process $\chi \bar\chi \to H^\dagger H$.\footnote{The rate of washout due to $\chi_\text{LH} \to \chi_\text{RH}$ mediated by a thermal loop of $H$ particles is of the same order, so at the level of our dimensional analysis it can be considered part of $\Gamma_5$.
Although the thermal loop is not exponentially suppressed at large $m_\chi(z)$, in the region where this suppression is significant washout is dominated by the vev-induced mass $m_\chi(z)$ (which is captured by $\Gamma^\chi_M$, see below).}
Similarly, we write the washout rate due to the CP-violating dimension-5 term in \cref{eq:LY} as\footnote{One may worry that also the dimension-4 Yukawa coupling $y_\chi \phi \, \overline{\chi_L} \chi_R$ might lead to washout of the $\chi$ asymmetry.
This would be problematic only in the false-vacuum phase, as in the true-vacuum phase the large mass of $\chi$ leads to complete washout anyway.
However, note that chirality-changing $2 \to 1$ processes like $\chi_\text{LH} \bar\chi_\text{LH} \to \phi$ are kinematically forbidden due to the thermal mass of $\chi$.
The $2 \to 2$ process $\chi \bar\chi \to \phi\phi$ mediated by $t$-channel $\chi$ exchange, on the other hand, always annihilates pairs of opposite-helicity $\chi$ particles and antiparticles for massless $\chi$.
The relevant thermal loop contributions to the fermion self-energy, usually referred to as thermal masses, do not invalidate the last argument because, unlike regular mass terms, they do not change chirality.}
\begin{align}
    \tilde\Gamma_5 
    &= T\ba{y_\chi^2 T_I^\chi\frac{T}{m_\chi}}^{\!2} 
    {\exp}\ba{{-}\frac{m_\chi(z)}{T}} \,.
\end{align}
Here again Boltzmann suppression freezes in the $\chi$ asymmetry in the true-vacuum phase.
Both $\Gamma_5$ and $\tilde\Gamma_5$ correspond to processes that wash out the asymmetry in $\chi$, and they do so without creating a corresponding scalar asymmetry.
($H$, $H^\dagger$ are produced in equal numbers by $\chi \bar\chi \to H^\dagger H$, and for the real scalar $\phi$, particle and antiparticle are identical.)

A third washout process is mixing between left- and right-chiral $\chi$ fields via the vev-induced mass term.
The rate for this process is described by the relaxation rate $\Gamma^\chi_M$, which can be calculated following refs.~\cite{Fuchs:2020pun, deVries:2017ncy, Lee:2004we, Cirigliano:2006wh}.
$\Gamma^\chi_M$ vanishes in the symmetric phase ($z \ll 0$), but is the dominant washout process in the broken phase ($0 \ll z$).
For the dimension-6 $N_R$ portal, the rate $\Gamma_Y^N$ also enters, see \cref{sec:app-inputs}.

Finally, the electroweak sphaleron conversion rate, $\Gamma_\textrm{ws}$, can be found in ref.~\cite{Fuchs:2020pun} (see also ref.~\cite{Bodeker:1999gx}).
Taking these processes together, we see how a chiral asymmetry in $\chi$ is transmitted first to an asymmetry in $\ell$ via the portal interactions, and then to an asymmetry in $q$ via the electroweak sphalerons.

The $\tilde\mu_a$ factors for the model with the dimension-6 $N_R$ portal operator, \cref{eq:portal-NR}, are defined as
\begin{align}
    \begin{split}
	    \tilde\mu_{\rm p}   &\equiv - 2\frac{\chi}{k_\chi} - 2\frac{N}{k_N}\,, \\
	    \tilde\mu_{\rm 5}   &\equiv - 2\frac{\chi}{k_\chi}\,,\\
	    \tilde\mu^\chi_M    &\equiv - 2\frac{\chi}{k_\chi}  \,,\\
	    \tilde\mu_Y^N   &\equiv \frac{N}{k_N} - \frac{\ell}{k_\ell} - \frac{h}{k_h}\,, \\
        \tilde\mu_\text{ws} &\equiv 3\frac{\ell}{k_\ell} + 9\frac{q}{k_q}\,,
    \end{split}
    \label{eq:mu-tilde-dim-6}
\end{align}
while those for the model with the dimension-8 Weinberg portal operator, \cref{eq:portal-W}, are
\begin{align}
    \begin{split}
	    \tilde\mu_{\rm p}   &\equiv - 2\frac{\chi}{k_\chi} - 2\frac{\ell}{k_\ell} - 2\frac{h}{k_h}\,, \\
	    \tilde\mu_{\rm 5}   &\equiv - 2\frac{\chi}{k_\chi}\,,\\
	    \tilde\mu^\chi_M    &\equiv - 2\frac{\chi}{k_\chi}  \,, \\
        \tilde\mu_\text{ws} &\equiv 3\frac{\ell}{k_\ell} + 9\frac{q}{k_q}\,.
    \end{split}
    \label{eq:mu-tilde}
\end{align}
The right-hand sides of \cref{eq:mu-tilde-dim-6,eq:mu-tilde} consist of linear combinations of the ``effective chemical potentials'' of the particle species involved in the respective processes.
The term ``effective'' here refers to the fact that a factor $T^2/6$ has been absorbed into the definition of the chemical potentials.
More precisely, the effective chemical potential $\tilde\mu_a$ of a species $a = \chi, N, \ell, h, q$ is related to its physical chemical potential, $\mu_a$, and to the particle number asymmetry, $a$, via
\begin{align}
    a = n_a - n_{\bar a}
        &= \frac{\mu_a T^2}{6} \, k_a(m_a/T) + \cO(\mu_a^3)  \notag\\
        &= \tilde\mu_a \, k_a(m_a/T) + \cO(\tilde\mu_a^3) \,.
    \label{eq:n-mu}
\end{align}
Explicit expressions for the  mass-dependent factors $k_a(m_a/T)$ can be found, for instance, in refs.~\cite{Fuchs:2020pun, deVries:2017ncy}. \Cref{eq:n-mu} is obtained by expanding $n_a$, when written in terms of a momentum-integral over a Fermi--Dirac (or Bose--Einstein) distribution, to linear order in $\mu_a$.  
For massless particles, $k_a(0)$ is simply the number of (Weyl fermion or real scalar) degrees of freedom; in particular $k_\chi(0) = k_N(0) = 1$, $k_\ell(0) = 2$, $k_h(0) = 4$, and $k_q(0) = 6$. 
Since our mechanism acts prior to the electroweak phase transition, the $k_a$ of the Standard Model particles are calculated based on their thermal masses only, where the latter can be found, for instance, in ref.~\cite{Fuchs:2020pun}.
We find that for the fermions there is almost no change from the massless approximation, but $k_h$ changes to $\approx 3$.
We evaluate $k_\chi$ at its $z$-dependent mass and find that $k_\chi \approx 1$ when $\chi$ is approximately massless and $k_\chi \approx 0$ when the $\chi$ mass is much larger than the temperature.

The integer factors appearing in \cref{eq:mu-tilde-dim-6,eq:mu-tilde} can be understood from the linear expansion in the chemical potentials of the interaction rates.  
For example, the rate of the process induced by the dimension-8 Weinberg portal operator in \cref{eq:portal-W}, $\chi\bar \chi \to 2 L + 2 H$, contains $f_\chi$, $f_{\bar\chi}$, $f_\ell^2$ and $f_h^2$, where $f_a$ is the phase space distribution function of species $a$.  
When expanded to leading order, we obtain the factors of 2 seen in $\tilde \mu_{\rm p}$ in \cref{eq:mu-tilde}. 
The integer coefficients in front of the $\Gamma \tilde\mu$ terms in \cref{eq:transport-eq-dim-6,eq:transport-eq-dim-8,} reflect the number of particles affected by each process. 
That is, since the process $\chi\bar \chi \to 2 L + 2 H$ occurs for each lepton generation, three $\chi$ particles and three $\bar\chi$ particles can annihilate to create two first-generation leptons, two second-generation leptons, and two third-generation leptons, along with six Higgs particles.  
Since $\ell$ counts the asymmetry \emph{per generation}, $\ell$ increases by $2\Gamma_{\rm p} \tilde\mu_{\rm p}$ and $h$ by $6\Gamma_{\rm p} \tilde\mu_{\rm p}$, while $\chi$ and $-\bar\chi$ both reduce by $3\Gamma_{\rm p} \tilde\mu_{\rm p}$.

The chiral asymmetry in $\chi$ arises from CP-violating interactions between the $\chi$ particles and the bubble wall, which are encoded in the source terms $S_\chi$ in \cref{eq:transport-eq-dim-6,eq:transport-eq-dim-8}.  
For the wall profile we assume
\begin{align}
    v_\phi(z) &= \frac{v_\phi}{2}
                 \bb{1 + {\tanh}\ba{\frac{3z}{l_w}}} \,,
    \label{eq:wall-profile}
\end{align}
where $l_w$ parameterises the wall width.  
When $v_\phi(z) \neq 0$, the CP-violating coupling between $\phi^2$ and $\overline{\chi}_L \chi_R$ in \cref{eq:LY} gives rise to a complex phase in the mass of $\chi$,
\begin{align}
	\overline{m}_\chi(z) &= y_\chi v_\phi(z)
	    \ba{1 - i\, T^\chi_I \frac{v_\phi(z)}{v_\phi}} \,.
	\label{eq:mbar}
\end{align}
Throughout our paper we will use the shorthand notations $m_\chi(z) \equiv |\overline{m}_\chi(z)|$ and $\theta(z)\equiv\text{arg}(\overline{m}_\chi(z))$. 
The CP-violating source term $S_\chi$ in \cref{eq:transport-eq-dim-6,eq:transport-eq-dim-8} can be written in a semi-classical approximation as~\cite{Cline:2000kb,Cline:2020jre,Kainulainen:2021oqs} (see also \cite{Cline:1997vk, Cline:2000nw, Cline:2001rk, Kainulainen:2001cn, Kainulainen:2002th, Kainulainen:2021zbf})
\begin{align}
     S_\chi(z) &=
      - \frac{S_1}{D_0} + \frac{v_w S^\prime_1 + S^\prime_2}{D_0\Gamma_{\rm tot}} 
     \xrightarrow[v_w\ll 1]{}
     \frac{S^\prime_2}{K_1\Gamma_{\rm tot}} \,,
     \label{eq:baryo:source}
\end{align}
where $D_0=K_1$ in the small-$v_w$ limit,  $\Gamma_{\rm tot}=K_4/(D_{\chi} K_1) $ is the total interaction rate, the $z$-dependent diffusion coefficient of $\chi$ is given in \cref{sec:app-diff-coeff}, 
\begin{align}
    % S_2(z) &= v_w\frac{T^2}{6}\left[{-}K_8(z)\left(m_\chi^2(z)\theta^\prime(z)\right)^\prime + K_9(z) \, m_\chi^2(z) \left(m_\chi^2(z)\right)^\prime \theta^\prime(z) \right]\,, 
    % \label{eq:baryo:source}
    S_2^\prime &= v_w\frac{\partial}{\partial z}\!
    \left[{-}K_8(z)\left(m_\chi^2(z)\theta^\prime(z)\right)^\prime + K_9(z) \, m_\chi^2(z) \left(m_\chi^2(z)\right)^\prime \theta^\prime(z) \right]\,, 
\end{align}
and the derivatives are with respect to $z$, and~\cite{Fromme:2006wx}
\begin{align}
    K_1(z) &= -\left\langle\frac{p_z^2}{E}\frac{\partial^2 f_\chi}{\partial E^2}\right\rangle\,,\nonumber\\
    K_4(z) &= \left\langle\frac{p_z^2}{E^2}\frac{\partial f_\chi}{\partial E}\right\rangle\,,\nonumber\\
    K_8(z) &= \left\langle\frac{\abs{p_z}}{2E^2 E_z}\frac{\partial f_\chi}{\partial E}\right\rangle\,,\nonumber\\
    K_9(z) &= \left\langle\frac{\abs{p_z}}{4E^3 E_z}\left(\frac{1}{E}\frac{\partial f_\chi}{\partial E}-\frac{\partial^2 f_\chi}{\partial E^2}\right)\right\rangle\,.
\end{align}
Here, $f_\chi$ is the Fermi--Dirac distribution for $\chi$, $E$ is the total energy of the particle, $E = (m_\chi^2(z) + p_x^2 + p_y^2 + p_z^2)^{1/2}$, and $E_z \equiv (m_\chi^2(z) + p_z^2)^{1/2}$. The average is defined as
\begin{align}
	\vev{X} = \frac{\int\upd^3p \,X(\vec{p})}{\int\upd^3p \,\frac{\partial f_\chi(m_\chi=0)}{\partial E}}\,.
	\label{eq:averaging}
\end{align}
We have now defined all the quantities appearing in \cref{eq:transport-eq-dim-6,eq:transport-eq-dim-8}.
They are again summarised in \cref{tab:inputs} in \cref{sec:app-inputs}.

Note that there is an ongoing debate in the literature on electroweak baryogenesis regarding the most appropriate method for calculating the source term, with the main contenders being the semiclassical, or WKB, approach, and the vev-insertion approximation~(VIA), in which the mass of the particle that develops a chiral asymmetry is treated perturbatively~\cite{Cline:2021dkf}.
For electroweak baryogenesis, calculations based on the VIA source predict a final baryon asymmetry several orders of magnitude larger than those based on a WKB source term.
The reason is that the VIA source has the same sign for all $z$, while the WKB source features both positive and negative contributions that largely cancel each other when integrated over the width of the bubble wall.
For our purposes, the VIA would not be applicable in the first place because the mass of $\chi$ cannot be treated as a small perturbation.
This is the main reason we have chosen the semiclassical approach, but we also note that this approach typically leads to a more conservative estimate of the asymmetry generated due to cancellations between positive and negative contributions to the source term.

To solve the transport equations given in \cref{eq:transport-eq-dim-6,eq:transport-eq-dim-8}, we first rewrite the four (five) second-order differential equations in terms of eight (ten) first-order differential equations.
We solve these equations numerically with the boundary conditions $a(z \to -\infty) = 0$ and $a'(z \to \infty) = 0$. 
The first boundary condition ensures vanishing asymmetries deep in the false-vacuum phase, while the second one stipulates that the asymmetries must have settled down to a constant (but possibly non-zero) value deep inside the true-vacuum phase. 
As $D_\chi \gg l_w$ in the broken phase, we are dealing with a problem that involves two very different length scales. 
To improve the numerical stability, we solve the equations in terms of the variable $\tilde z = \arsinh(3z/l_w)$, which depends linearly on $z$ near the bubble wall (of width $l_w$) and logarithmically on $z$ far from the wall.
In simplified situations, where all $z$-dependent quantities can be described as step functions, the system is analytically solvable \cite{Fuchs:2020pun}. We have verified that our numeric results coincide with the analytic solutions in applicable cases.

%-----------------------------------------------------------------------------
\section{Results}
\label{sec:results}
%-----------------------------------------------------------------------------

We are now ready to present and discuss our numerical solutions to the transport equations, \cref{eq:transport-eq-dim-6,eq:transport-eq-dim-8}.

%-----------------------------------------------------------------------------
\subsection{Benchmark Points}
\label{sec:benchmark-points}
%-----------------------------------------------------------------------------

\begin{figure}[t]
	\centering
    \includegraphics{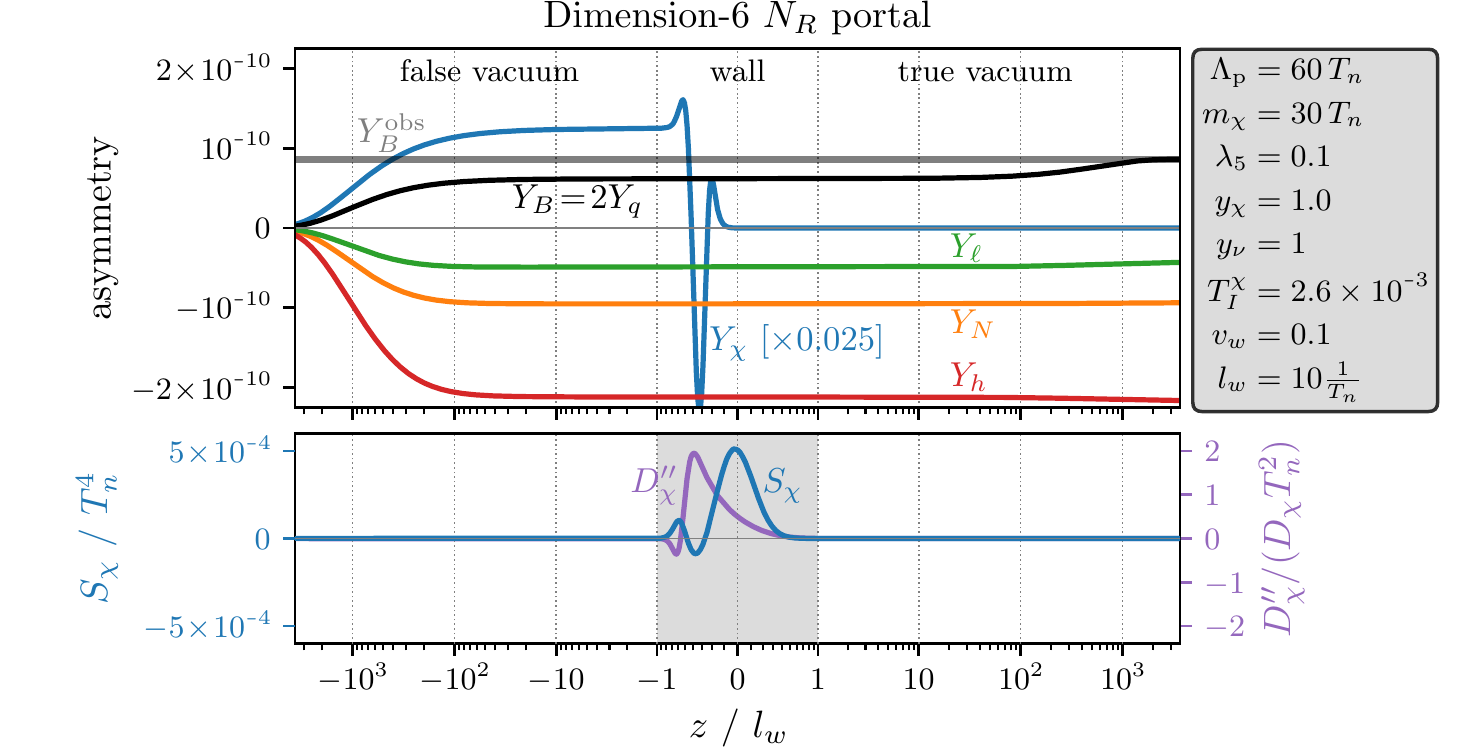} \\[0.5cm]
    \includegraphics{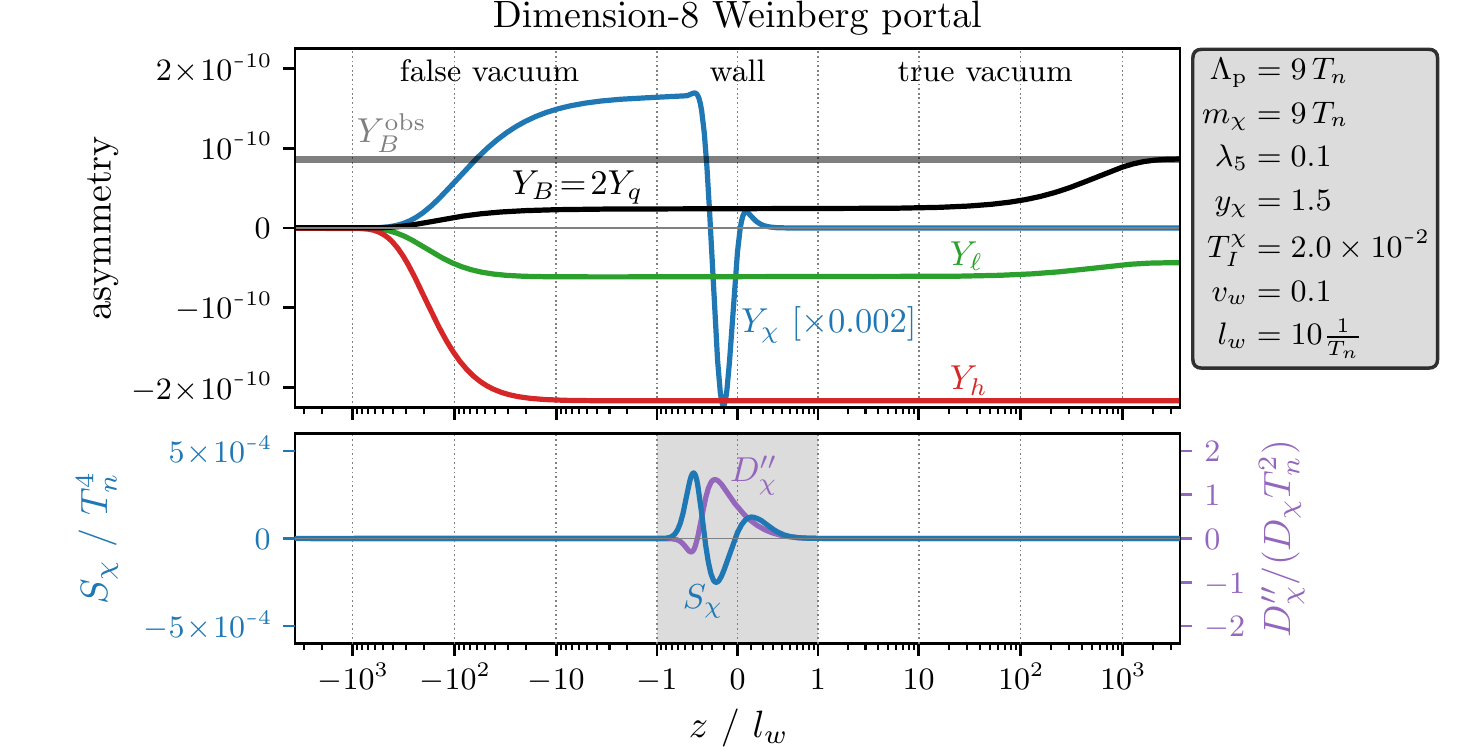} 
	\caption{Solution to the transport equations in the stationary state for two representative parameter points, one using the dimension-6 $N_R$ portal (top), and one using the \mbox{dimension-8} Weinberg portal (bottom). 
	The coloured curves show how the particle--antiparticle asymmetries of $\chi$ (blue) and different SM particle species (orange, green, red, black) are distributed in space. The black curves labelled $Y_B$ show the final baryon asymmetry, and the horizontal grey band indicates the observed value of $Y_B$. The vertical grey band illustrates the width of the bubble wall.
	Below each panel we show the source term $S_\chi$ (blue) and the term proportional to $D''$ in the Fokker--Planck equation.
	Note that the horizontal axis is linear in the variable $\tilde z = \arsinh(3z/l_w)$, whose dependence on $z$ is approximately linear close to the wall ($|\tilde{z}| \lesssim 1$), and logarithmic far away from it ($|\tilde{z}| \gg 1$).}
	\label{fig:solution}
\end{figure}

We begin in \cref{fig:solution} by showing the $z$-dependence of the various asymmetries in the stationary state for two representative parameter points, one for the dimension-6 $N_R$ portal (top) and one for the dimension-8 Weinberg portal (bottom).  
All asymmetries are normalised to the entropy density of the Universe. 
That is for each particle species $a$ we show the quantity
\begin{align}
    Y_a \equiv \frac{n_a - n_{\bar{a}}}{s} \,,
    \label{eq:Ya}
\end{align}
where $n_a$ and $n_{\bar{a}}$ are the number densities of $a$ particles and anti-particles, respectively, while $s = 2 \pi^2 g_\star T_n^3/45$ is the cosmological entropy density and $g_\star$ is the effective number of relativistic degrees of freedom in the thermal bath.
The advantage of using $Y_a$ as a variable is that $Y_a$ remains constant  from the end of baryogenesis until the present day.
Note that the horizontal axes in \cref{fig:solution} are linear in the variable $\tilde z = \arsinh(3z/l_w)$, whose dependence on $z$ is approximately linear close to the wall and logarithmic far away from it.
This means that the plots can clearly illustrate the fast processes occurring close to the wall, while simultaneously showing features due to slower processes, such as electroweak sphalerons.

We see that in both cases a spike in the chiral asymmetry of $\chi$ (blue curve) is generated close to the wall ($z \approx 0$).
This spike in the asymmetry is partly caused by the source term $S_\chi$ (blue curves in the bottom sub-panels of \cref{fig:solution}), which is zero far in front of the wall, then becomes positive as the wall approaches before flipping sign twice due to the interplay of the two terms in \cref{eq:baryo:source}, and returning to zero at $z \gg 0$.
As mentioned in \cref{sec:transeqs}, the occurrence of both positive and negative contributions is a well-known feature of source terms derived in the semi-classical approximation, as opposed to expressions obtained from the vev-insertion approximation, which do not flip sign~\cite{Cline:2021dkf}.
Close to the wall, the term proportional to $D''_\chi(z)$ in the Fokker--Planck operator, \cref{eq:fokker-planck}, also plays an important role.
In both cases, this term leads to a net flow of asymmetry to the $z < 0$ region.

Once the asymmetry is produced, it diffuses away from the wall into both the false vacuum phase and the true vacuum phase.
However, the asymmetry cannot survive in the true vacuum phase since it is quickly washed out by the $\chi$ mass term ($m_\chi(z \gtrsim 0)\gtrsim T_n$).  In the transport equations, \cref{eq:transport-eq-dim-6,eq:transport-eq-dim-8}, this is accounted for by the term containing $\Gamma_M^\chi$, which mixes left-handed and right-handed $\chi$ particles.
In the false vacuum phase, $\chi$ particles suffer some washout from the $\Gamma_M^\chi$ term, but in both the dimension-6 and dimension-8 cases retain a net negative asymmetry far in front of the wall.

Deep inside the false vacuum phase ($z \lesssim -100\,l_w$ for the chosen dimension-6 portal benchmark point, $z \lesssim -\text{few} \times 10\,l_w$ for dimension-8), the portal interaction converts part of this asymmetry into an asymmetry in $\ell$ and $h$.
The exact distance at which this happens is different for the two benchmark points mostly because of the different choices for $y_\chi$, which lead to different $\chi$ diffusion coefficients.
For the dimension-6 $N_R$ portal, an asymmetry in $N_R$ is produced, but thanks to the assumption of a large neutrino Yukawa coupling $y_\nu$, the $N_R$ asymmetry quasi-instantaneously spreads to the SM lepton doublets, $\ell$, and Higgs bosons, $h$.
For the dimension-8 Weinberg portal, the asymmetry in $\chi$ is directly converted into $\ell$ and $h$ asymmetries.

After a time $1/\Gamma_\text{ws}\sim \num{e5}/T_n$ the SM weak sphalerons convert part of the $\ell$ asymmetry into a baryon asymmetry ($B$), see \cref{sec:app-inputs}.
This corresponds to an increase of $B$ on distance scales around $\pm 1000 \, l_w$.
The change in $Y_B$ due to sphaleron processes is observed both in the false-vacuum phase and in the true-vacuum phase, even though the portal interaction that seeds a lepton asymmetry is active only in the false vacuum.
This is because the advancing bubble wall does not give $Y_\ell$ and $Y_q$ time to equilibrate before being swept into the true-vacuum phase, so a significant lepton asymmetry passes into the true vacuum before coming into equilibrium with $Y_q$.
This equilibration then happens in the true vacuum phase, as long as sphalerons are active in both phases.
This is the case here, but would not be if the dark sector and electroweak phase transitions were one and the same.

For the parameter points shown in \cref{fig:solution} the final baryon asymmetry matches the observed asymmetry in the Universe \cite{Planck:2018vyg},
\begin{align}
    Y_B^\text{obs} = \num[separate-uncertainty=true]{8.65(4)e-11}\,,
    \label{eq:YB-obs}
\end{align}
as indicated by the horizontal grey band.
It is interesting to note the different scale of the $\chi$ asymmetry and the SM asymmetries.
For example, in the dimension-8 case at $z = -l_w$, a $\chi$ asymmetry $Y_\chi \approx \num{8e-8}$ generates a lepton asymmetry $Y_\ell \approx \num{-6e-11}$ and a baryon asymmetry $Y_B \approx \num{2e-11}$.
Only a fraction of the asymmetry is converted since the portal operators are suppressed by a relatively large scale, $\Lambda_\text{p}$.

Note that the parameter points chosen in \cref{fig:solution} have $T_n \ll \Lambda_\text{p}$ but $\Lambda_\text{p}$ of the same order of magnitude as $m_\chi$.
One might wonder to what extent the EFT approach used for the portal operators and the dimension-5 operator, \cref{eq:L5} is valid in this case.
However, in the false vacuum phase, where these operators are relevant, $\chi$ is still massless and so the conditions for using EFT are very well satisfied in the region where EFT is needed.

%-----------------------------------------------------------------------------
\subsection{Parameter Scans}
\label{sec:param-scans}
%-----------------------------------------------------------------------------

\begin{figure}[p]
    \vspace*{-0.4cm}
	\centering
	Dimension-6 $N_R$ portal\\[0.2cm]
    \includegraphics{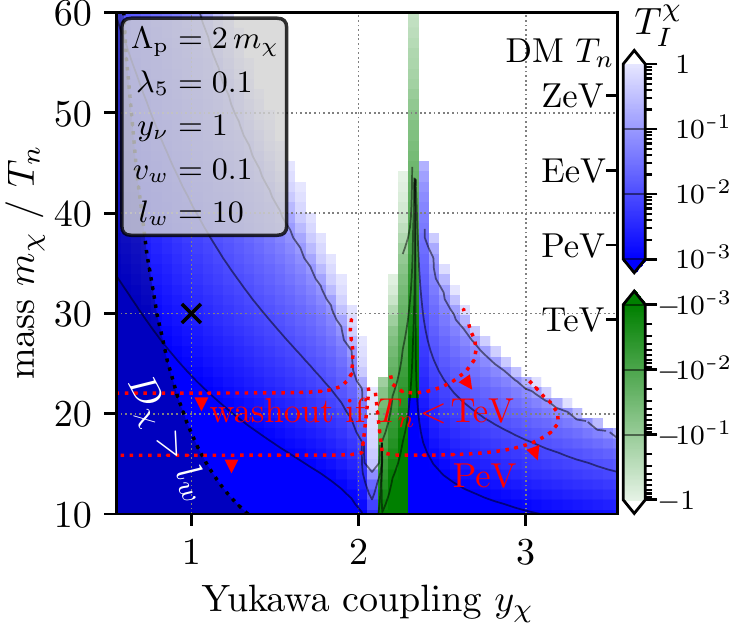}  \hfill
    \includegraphics{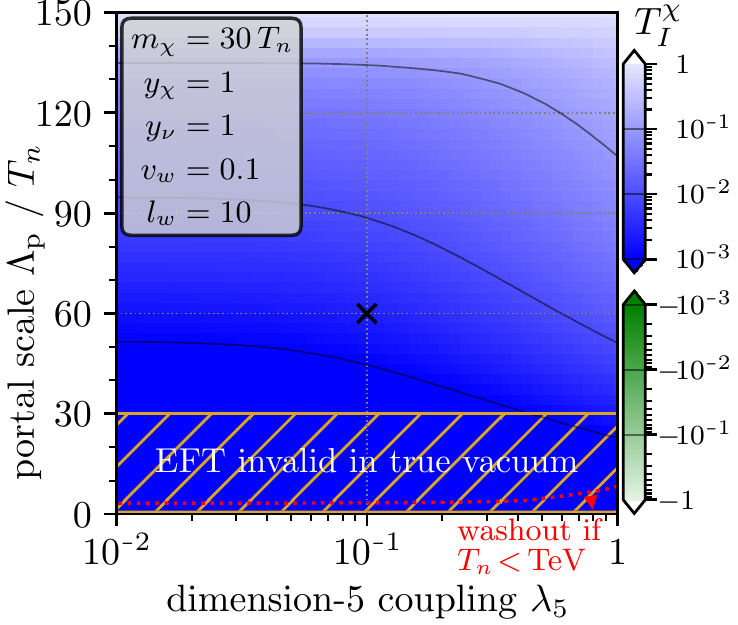} \\[0.3cm]
    Dimension-8 Weinberg portal\\[0.2cm]
    \includegraphics{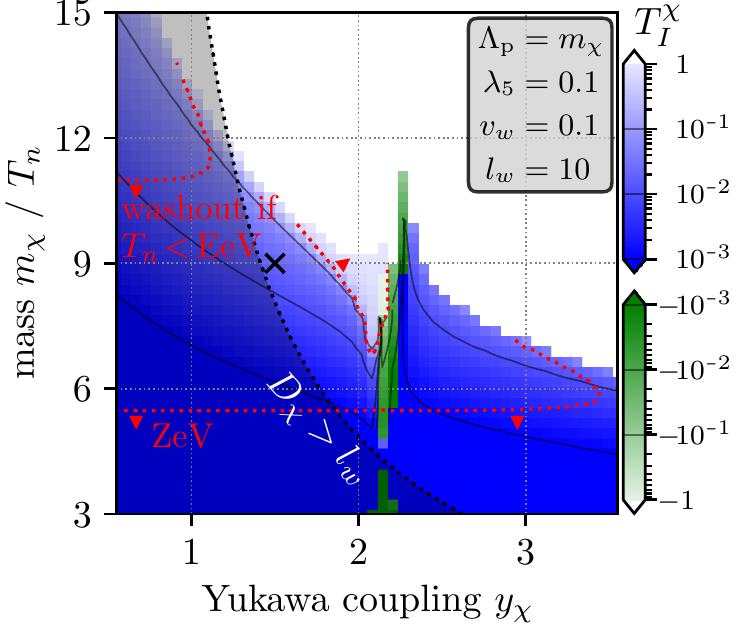}  \hfill
    \includegraphics{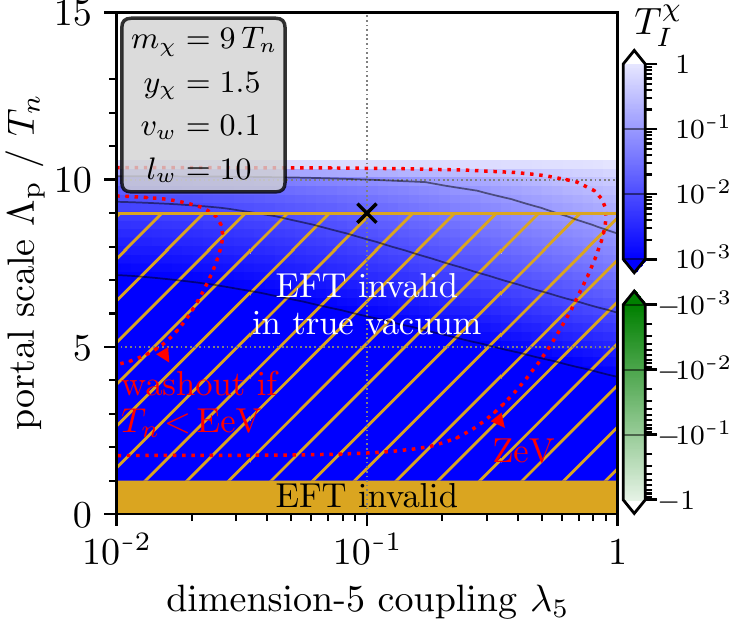} 
	\caption{Parameter space of Filtered Baryogenesis with the dimension-6 $N_R$ portal (top) and with the dimension-8 Weinberg portal (bottom). 
	The colour code and the black contours indicate the amount of CP-violation (parameterised as $T_I^\chi$, one black contour per order of magnitude) required to reproduce the observed baryon asymmetry in the Universe.
	In the white regions, $|T_I^\chi|>1$ would be required, invalidating our EFT treatment of the CP-violating interaction.
	In the shaded grey region, the WKB approximation used in the source term is not valid because $D_\chi(z=0) > l_w$.
	We exclude the solid orange region in the bottom right panel ($\Lambda_\text{p} < T_n$) because an EFT description of the portal interaction is not valid there.
	In addition, the orange hatched region indicates where $T_n < \Lambda_p < m_\chi$, that is, where EFT is valid in the false vacuum, but not in the true vacuum.
	Red dotted lines indicate the lower limits on the phase transition temperature, $T_n$, to avoid washout of the asymmetry after the dark sector phase transition and before the electroweak one.
	We show this washout limit only where it is above the electroweak scale, which constitutes a general lower limit for our mechanism.
	Black crosses represent the parameter points used in \cref{fig:solution}.
	In the top-left panel, we also indicate the nucleation temperature, $T_n$, needed to set the correct dark matter abundance at a given value of $m_\chi$ via the Filtered Dark Matter mechanism.
	}
	\label{fig:paramspace}
\end{figure}

A broader exploration of the parameter space is shown in \cref{fig:paramspace}, again for both the dimension-6 $N_R$ portal (top row) and for the dimension-8 Weinberg portal (bottom row).  
In these plots, the colour code and the black contours indicate the sign and size of the CP-violating phase, parameterised by $T_I^\chi$, that is required to reproduce the observed baryon asymmetry.
The white regions would require $|T_I^\chi|>1$, implying a breakdown of the EFT for the CP-violating interactions.
Similarly, in the solid orange region, where $\Lambda_\text{p} < T_n$, an EFT description of the portal interaction ceases to be valid.
Finally, in the orange hatched region, $T_n < \Lambda_p < m_\chi$, that is EFT is valid in the false vacuum, where the portal is actually active, but not in the true vacuum, where the portal interaction is irrelevant anyway.
Therefore, in that region our results should still be reliable even though EFT is not strictly applicable.

The left hand panels reveal a strong dependence on $m_\chi$. 
This is expected since the chiral asymmetry in $\chi$ is mostly generated in momentum modes with $p \sim m_\chi$.  That is, modes that are on the threshold of being able to enter the true vacuum phase. 
For $m_\chi\gg T_n$, these modes are located further out in the tails of the thermal distribution,  and so they are more Boltzmann suppressed.
This suppression needs to be compensated by initially producing a larger asymmetry, which in turn requires more CP-violation.
Furthermore, washout by the mass term is more efficient at larger $m_\chi$.
The required amount of CP-violation also depends strongly on $y_\chi$ and even involves sign flips.
This is due to the intricate interplay between the source term $S_\chi(z)$ and the $D_\chi''(z)$ term in the transport equations, together with the overall size of $D_\chi(z)$ (which depends on $y_\chi$). 

The right-hand panels of \cref{fig:paramspace} show a strong dependence on the portal scale $\Lambda_\text{p}$.
This is not unexpected since a higher portal scale corresponds to a weaker portal coupling, so less of the $\chi$ asymmetry is transferred to the SM.
Above $\Lambda_\text{p} \approx 150 T_n$ (top) and $\Lambda_\text{p} \approx 10 T_n$ (bottom), $T_I^\chi$ would need to be larger than 1 to give the observed baryon asymmetry.
We also notice an interesting dependence on $\lambda_5$: below a certain threshold, this parameter is irrelevant to the final baryon asymmetry; above the threshold, it reduces the chiral asymmetry of $\chi$ and a stronger CP-violation is required to compensate.

Note that we have kept the bubble wall velocity, $v_w$, and width, $l_w$, fixed in \cref{fig:paramspace}.
We find that our results are only weakly dependent on these parameters.

Besides the EFT conditions discussed above, there are three situations where the formalism in this paper ceases to be valid.
Firstly, if $D_\chi(z=0) > l_w$, the source term $S_\chi$ in the transport can no longer be written using the WKB approximation as in \cref{eq:baryo:source}.
The region where $D_\chi(z=0) > l_w$ is located to the left of the black dotted curves in the left-hand panels of \cref{fig:paramspace}.
These curves should not be understood as hard limits, and certainly not as a constraint on the viability of the Filtered Baryogenesis mechanism.
Rather, they indicate where our approximations start to introduce larger errors into our quantitative results, while qualitatively, our calculations may still hold.

Secondly, when the diffusion coefficient $D_\chi$ is large, the mean free path may no longer be shorter than the other length scales in the problem, and the diffusion ansatz for the transport equations may no longer be justified.
Instead, it would then be more appropriate to describe the dynamics of $\chi$ particles in a formalism similar to the one developed in \cite{Baker:2019ndr, Baker:2021nyl, Baker:2021sno}, where the trajectories of individual momentum modes are tracked through their interactions with the wall, to determine which $\chi$ particles are able to enter the true vacuum phase and which ones are reflected back into the false vacuum.
We estimate in \cref{sec:app-diff-approx} how such a treatment would impact our results and conclude that the viability of the mechanism remains unaffected.

Thirdly, the portal rate plus the dimension-5 washout rate, which are suppressed but non-zero in the true vacuum, may lead to a depletion of the baryon asymmetry long after the wall has passed.
This would be the case if $\Gamma_\text{p}+\Gamma_5+\tilde\Gamma_5 > H$, which corresponds to the region below the dotted red lines for different transition temperatures $T_n$.

\FloatBarrier

%-----------------------------------------------------------------------------
\subsection{Relation to the Filtered Dark Matter Mechanism, and Direct Detection Constraints}
\label{sec:dd}
%-----------------------------------------------------------------------------

\begin{figure}[t]
	\centering
	\includegraphics{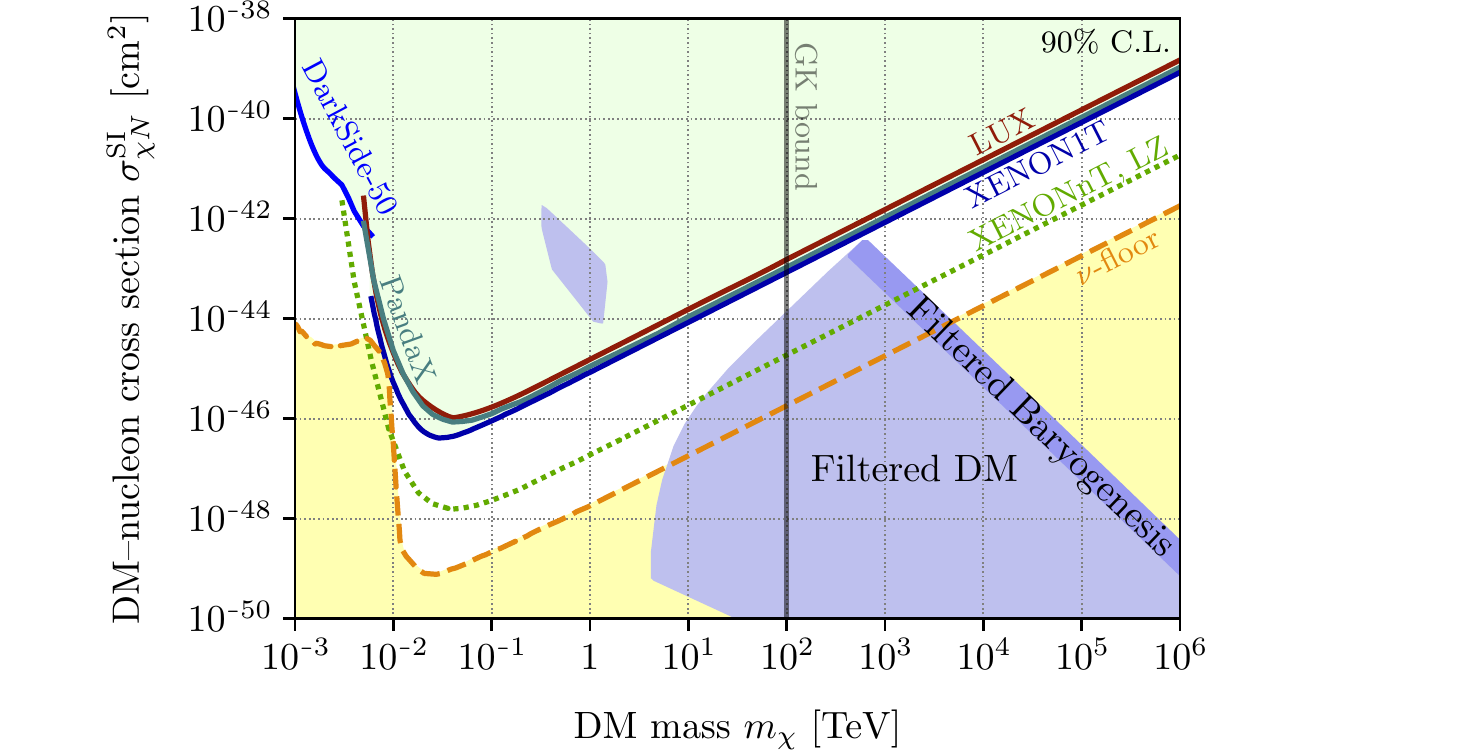}
	\caption{The spin-independent dark matter--nucleon scattering cross section for Filtered Dark Matter and Filtered Baryogenesis. The faint purple region is similar to the one shown in ref.~\cite{Baker:2019ndr}, while the dark purple band indicates the region where the dark matter abundance and the baryon asymmetry of the Universe can be explained simultaneously using the formalism developed in the present paper.  We compare to current experimental limits (solid) and future sensitivities (dotted), as well as the neutrino floor (orange dashed line / yellow shaded region) \cite{Gutlein:2010tq, Kopp:2012dz, Billard:2013qya}. The vertical grey line indicates the Griest--Kamionkowski bound \cite{Griest:1989wd} that sets an upper limit on the mass of conventional WIMPs (but not on Filtered Dark Matter).}
	\label{fig:xsec}
\end{figure}

We finally come back to the original motivation for this work: explaining baryogenesis and the dark matter abundance in the Universe at the same time, building on the Filtered Dark Matter scenario from ref.~\cite{Baker:2019ndr}.
We show in \cref{fig:xsec} how the parameter regions in which Filtered Dark Matter can explain the observed dark matter abundance compare to the parameter regions in which we have demonstrated successful baryogenesis in the case of the dimension-6 $N_R$ portal.
In the case of the dimension-8 Weinberg portal, the values of $m_\chi / T_n$ which lead to successful baryogenesis are not large enough for the Filtered Dark Matter mechanism to yield the observed abundance of dark matter.
We present this in the $m_\chi$--$\sigma_{\chi N}^\text{SI}$ plane, where $\sigma_{\chi N}^\text{SI}$ is the cross section for spin-independent dark matter--nucleon scattering which is probed by direct detection experiments, see \cref{eq:sigma-dd}.
As already noted in ref.~\cite{Baker:2019ndr}, the mechanism is particularly successful at large dark matter masses ($m_\chi \gtrsim \SI{10}{TeV}$), including large swaths of parameter space above the Griest--Kamionkowski (GK) bound \cite{Griest:1989wd}.
This bound precludes the existence of such heavy dark matter particles in conventional thermal freeze-out scenarios.
Note that there is a one-to-one correspondence between the dark matter mass and the phase transition temperature required to obtain the correct relic abundance \cite{Baker:2019ndr}.
As we choose $m_\phi \sim T_n$, this one-to-one correspondence also extends to $m_\phi$.

We first discuss the shape of the viable Filtered DM region (faint purple region in \cref{fig:xsec}).
The main features of this region are determined by the conditions that the dark sector Yukawa coupling $y_\chi$ and the Higgs portal coupling $\beta$ satisfy perturbative unitarity constraints ($y_\chi, \beta < \sqrt{4\pi}$), that $\chi$ is in equilibrium in the false vacuum but decoupled in the true vacuum, and that $\phi$ is in equilibrium with the SM sector in both phases.
Note that in ref.~\cite{Baker:2019ndr}, we assumed two species to be in equilibrium with each other when their interactions are faster than the Hubble rate.
Here, we apply the stricter criteria that the interaction rate should be faster than the time the phase transition takes to complete, which we ad-hoc take to be $1/(\num{1000} H)$~\cite{Apreda:2001us}.
This affects the lower edge of the purple region in \cref{fig:xsec}.

At low $m_\chi$ ($\lesssim \SI{100}{GeV}$), the temperature of the phase transition required to obtain the correct dark matter abundance needs to be well below the electroweak scale.
Keeping $\phi$ in equilibrium is then impossible because $\phi$--SM interactions are suppressed by a small $\phi$--Higgs mixing angle and by small fermion Yukawa couplings.
This problem is eased at dark matter masses between roughly \SI{100}{GeV} and \SI{1}{TeV} (corresponding to $T_n \gtrsim \text{few GeV}$), as SM particles with larger Yukawa couplings become accessible.
However, as $m_\chi$ is increased further, such that the corresponding $T_n$ and $m_\phi$ become $\gtrsim \text{few GeV}$, new SM final states cease to appear.
To keep $\phi$ in equilibrium, it is then necessary to increase the Higgs portal coupling $\beta$ and thereby the $\phi$--Higgs mixing angle.
Eventually, at $m_\chi \sim \text{several TeV}$, it becomes impossible to do so while maintaining a scalar mass matrix that has an eigenvalue at the observed Higgs mass of \SI{125}{GeV}.
This explains the gap between the two faint purple regions in \cref{fig:xsec}.
At even higher $m_\chi$, $T_n$, and $m_\phi$, direct $\phi \phi \leftrightarrow H H$ annihilation becomes kinematically allowed, rendering Filtered Dark Matter viable again.
The upper right-hand edge of the faint purple region is defined by the perturbativity constraint on $\beta$: at fixed $m_\chi$ (and thereby fixed $m_\phi$), the scattering cross section $\sigma_{\chi N}^\text{SI}$ depends only on $\beta$, and grows as $\beta^2$ according to \cref{eq:sigma-dd}.

Turning now to the prospects of baryogenesis in the Filtered Dark Matter scenario, the viable parameter region for the dimension-6 $N_R$ portal (shown in \cref{fig:paramspace}), with $y_\chi \gtrsim 1$, translates into the dark purple band in \cref{fig:xsec}.
Large $y_\chi$ implies sizeable $\chi$ annihilation cross-section, and since we require the equilibration rate of $\phi$ to be faster than the annihilation rate of $\chi$, this also implies large $\beta$.
Large $y_\chi$ and $\beta$, in turn, maximise $\sigma_{\chi N}^\text{SI}$ at given dark matter mass.
We see that to satisfy the Filtered DM constraints, the dark matter mass should be $\gtrsim \si{PeV}$.
For smaller $m_\chi$ or smaller $y_\chi$, we anticipate that successful baryogenesis is still possible.
However, as shown in \cref{fig:paramspace}, the formalism from \cref{sec:transeqs} is then no longer applicable because the diffusion coefficient $D_\chi$ becomes too large.

%-----------------------------------------------------------------------------
\section{Conclusions} 
\label{sec:conclusions}
%-----------------------------------------------------------------------------

In this paper, we have augmented the Filter Dark Matter mechanism from ref.~\cite{Baker:2019ndr} with CP violating couplings and portal interactions, to explain the baryon--antibaryon asymmetry of the Universe.
The bubble walls that separate the false-vacuum and true-vacuum phases during the first-order cosmological phase transition, and which set the dark matter abundance in this scenario, then have different transmissivities for left-chiral and right-chiral dark matter particles.
As well as reducing the overall dark matter abundance in the true vacuum phase (as in the vanilla Filtered Dark Matter scenario), the phase transition then also generates a chiral asymmetry in the dark sector.
Non-renormalisable portal interactions convert this chiral asymmetry into a SM lepton asymmetry, see \cref{eq:portal-NR,eq:portal-W} for the two exemplary portals.
The lepton asymmetry is eventually converted to a baryon asymmetry by electroweak sphaleron processes.

We have seen in \cref{fig:paramspace,fig:xsec} that the model succeeds in explaining both the dark matter abundance and the observed baryon asymmetry over wide ranges of parameter space.
It is likely that the mechanism's viability extends even beyond the parameter ranges demonstrated here -- notably into the regime of small dark matter Yukawa coupling $y_\chi$ -- but the diffusion-based formalism we have employed is unable to make reliable quantitative predictions in this case.
Instead, it would be necessary to track individual momentum modes of the dark matter population separately, as done for instance in refs.~\cite{Baker:2021nyl,Baker:2021sno}.
We leave this for future work.
Other questions we plan to address in the future include a survey of scalar sectors that lead to first-order phase transitions with the required properties (in particular a large order parameter), as well as investigating the possibility that the Filtered Dark Matter phase transition can be identified with the electroweak one.

%-----------------------------------------------------------------------------
\section*{Acknowledgements}
%-----------------------------------------------------------------------------

It is a great pleasure to thank Vincenzo Cirigliano, Valerie Domcke, Elina Fuchs, Christopher Lee, Andrew Long, G\'eraldine Servant, Yehonatan Viernik, Jorinde van de Vis, and Jordy de Vries for sharing their expertise on electroweak baryogenesis with us and explaining innumerable subtleties to us.
We are particularly grateful to Kimmo Kainulaninen for pointing out a mistake in the CP-violating source term of the transport equations in the first version of this paper that appeared on the arXiv. 
MJB, JK, LM, and YS would also like to congratulate MB, whose life underwent several first-order phase transitions during the course of this project (which were each much more important than the ones discussed here).
The work of MB, JK, and LM has been partly funded by the German Research Foundation (DFG) in the framework of the PRISMA+ Cluster of Excellence and by the European Research Council (ERC) under the European Union's Horizon 2020 research and innovation programme (grant agreement No.\ \texttt{637506}, ``$\nu$Directions''). MB and JK have moreover profited from DFG Grant No.~KO-4820/4-1.
The work of YS is supported by grants from the NSF-BSF (No.~2018683), the ISF (No.~482/20), the BSF (No.~2020300) and by the Azrieli foundation.
Finally, we acknowledge using the TikZ-Feynman package \cite{Ellis:2016jkw} to produce \cref{fig:illustration}.

%-----------------------------------------------------------------------------
\appendix
\section{Connection to Neutrino Masses}
\label{sec:app-inverse-seesaw}
%-----------------------------------------------------------------------------

In the scenario with the dimension-6 $N_R$ portal, we have explicitly introduced right-handed neutrino fields $N_R$ as dynamical degrees of freedom, but we have argued that the $N_R$ cannot have large Majorana masses as this would lead to washout of the generated baryon and lepton number asymmetries.
This implies that the model does not admit a type-I seesaw mechanism as an explanation for neutrino masses, at least not without significant fine-tuning.
Here, we argue that our $N_R$ fields could instead be part of an inverse seesaw scenario \cite{Mohapatra:1986ks,CentellesChulia:2020dfh}.
In the inverse seesaw, neutrino masses are generated by the following terms in the Lagrangian:
\begin{align}
    \mathcal{L} \supset  y_\nu \overline{L} \tilde{H} N_R
                       + m_D' \overline{N_R} S
                       + \frac{1}{2} \mu \overline{S^c} S \,,
\end{align}
where $S$ is an auxiliary singlet fermion that carries lepton number, $m_D'$ is a Dirac mass that can be sizeable (typically of order TeV), while the lepton number-violating coupling $\mu$ is much smaller, typically of order \SI{10}{eV}.
All fields are understood to carry flavour indices, and $y_\nu$, $m_D'$, and $\mu$ are understood to be matrices in flavour space.
The resulting neutrino mass matrix in the basis $(\nu_L, N_R^c, S)$ has the structure
\begin{align}
    \mathcal{M} =
    \begin{pmatrix}
        0     & m_D    & 0 \\
        m_D^T &  0     & m_D' \\
        0     & m_D'^T & \mu
    \end{pmatrix} \,,
    \label{eq:M-inverse-seesaw}
\end{align}
with $m_D \equiv y_\nu v_H / \sqrt{2}$, and $v_H = \SI{246}{GeV}$ is the SM Higgs vev.
Diagonalizing $\mathcal{M}$, one finds that the light neutrino masses are
\begin{align}
    m_\nu = m_D (m_D'^T)^{-1} \mu \, m_D'^{-1} m_D^T \,.
\end{align}
For the choice $y_\nu \sim \mathcal{O}(1)$ from the main part of this paper, $m_\nu$ is of the required magnitude (\SI{0.1}{eV}) if, for instance, $m_D' \sim \si{TeV}$ and $\mu \sim \SI{10}{eV}$.

To avoid washout of the generated baryon number asymmetries, we have to require that the washout rate due to the lepton-number violating $\mu \overline{S^c} S$ coupling remains small compared to the Hubble rate until either the temperature has dropped sufficiently far below the $N_R$ mass, thereby effectively decoupling $S$ from the Standard Model sector, or until electroweak sphalerons have frozen out.
For the generic parameter values indicated above, this is indeed the case: the washout rate at $T \gtrsim \si{TeV}$ is \cite{Fuchs:2020pun,Lee:2004we}
\begin{align}
    \Gamma_M^S \sim \frac{\mu^2}{T} \lesssim \SI{e-10}{eV} \,,
\end{align}
which is much smaller than the Hubble expansion rate, $H \sim \sqrt{g_*} T^2 / M_\text{Pl} \gtrsim \SI{e-3}{eV}$.

Note that the presence of $S$ in the inverse seesaw model will reduce the baryon asymmetry of the Universe by an $\mathcal{O}(1)$ factor compared to the results from \cref{sec:results}.
This is because, when $S$ equilibrate with $N_R$, the availability of extra lepton-number carrying degrees of freedom means that less baryon asymmetry will be generated by electroweak sphalerons.
This can easily be compensated by a minor adjustment of the CP-violating phase, as parameterised by $T_I^\chi$.

%-----------------------------------------------------------------------------
\section{Extended Version of the Transport Equations}
\label{sec:app-extended-transport}
%-----------------------------------------------------------------------------

In \cref{eq:transport-eq-dim-6,eq:transport-eq-dim-8}, we presented a minimal version of the transport equations describing the asymmetries near the wall, where we neglect all Yukawa couplings and assume fast strong sphaleron interactions. When taking the effect of the third generation Yukawas as well as the finite strong sphaleron rate into account, the transport equations for the dimension-8 Weinberg portal, for example, become
\begin{align}
\begin{split}
    \partial_\mu j_\chi^\mu
    &=  \Gamma_{\rm p} (2\tilde\mu_{\rm p}^{1,2}+\tilde\mu_{\rm p}^3) + (\Gamma_{\rm 5}+\tilde \Gamma_5) \tilde\mu_{\rm 5} + \Gamma^\chi_M \tilde\mu^{\chi}_M + S_{\chi} \,,\\[1ex]
    \partial_\mu j_\tau^\mu
    &= 	- \Gamma^\tau_Y \tilde\mu^{\tau}_Y \,,\\[1ex]
    \partial_\mu j_{\ell_{1,2}}^\mu 
    &= 	2\Gamma_{\rm p} \tilde\mu_{\rm p}^{1,2} - \Gamma_\text{ws} \tilde\mu_\text{ws}\,,\\[1ex]
    \partial_\mu j_{\ell_{3}}^\mu 
    &= 	2\Gamma_{\rm p} \tilde\mu_{\rm p}^3 - \Gamma_\text{ws} \tilde\mu_\text{ws} + \Gamma^\tau_Y \tilde\mu^{\tau}_Y\,,\\[1ex]
    \partial_\mu j_h^\mu
    &= 2\Gamma_{\rm p} (2\tilde\mu_{\rm p}^{1}+\tilde\mu_{\rm p}^3)  - \Gamma^\tau_Y \tilde\mu^{\tau}_Y +\Gamma^t_Y \tilde\mu^{t}_Y -\Gamma^b_Y \tilde\mu^{b}_Y \,,\\[1ex]
    \partial_\mu j_u^\mu
    &= \Gamma_\text{ss} \tilde\mu_\text{ss}\,,\\[1ex]
    \partial_\mu j_t^\mu
    &= 	\Gamma_\text{ss} \tilde\mu_\text{ss}- \Gamma^t_Y \tilde\mu^{t}_Y \,,\\[1ex]
    \partial_\mu j_b^\mu
    &= 	\Gamma_\text{ss} \tilde\mu_\text{ss}- \Gamma^b_Y \tilde\mu^{b}_Y \,,\\[1ex]
    \partial_\mu j_{q_{1,2}}^\mu 
    &= -3\Gamma_\text{ws} \tilde\mu_\text{ws} -2\Gamma_\text{ss} \tilde\mu_\text{ss} \,,\\[1ex]
    \partial_\mu j_{q_{3}}^\mu 
    &= - 3\Gamma_\text{ws} \tilde\mu_\text{ws} -2\Gamma_\text{ss}\tilde\mu_\text{ss} + \Gamma^t_Y \tilde\mu^{t}_Y + \Gamma^b_Y \tilde\mu^{b}_Y \,,
\end{split}
\end{align}
with
\begin{align}
\begin{split}
	\tilde\mu_{\rm p}^{i} &= - 2\frac{\chi}{k_\chi} - 2\frac{\ell_{i}}{k_\ell} - 2\frac{h}{k_h}\,,\qquad \tilde\mu_{\rm 5} = - 2\frac{\chi}{k_\chi}\,,\qquad\tilde\mu^\chi_M = - 2\frac{\chi}{k_\chi}  \,,\\
	\tilde\mu^\tau_Y &=  \frac{\tau}{k_\tau} - \frac{\ell_3}{k_\ell}  + \frac{h}{k_h}\, ,\qquad \tilde\mu^t_Y = \frac{t}{k_t} - \frac{q_3}{k_q} - \frac{h}{k_h}\, ,\qquad \tilde\mu^b_Y = \frac{b}{k_b} - \frac{q_3}{k_q} + \frac{h}{k_h}\,,\\
     \tilde\mu_\text{ws} &= 2\frac{\ell_{1,2}}{k_\ell}+\frac{\ell_3}{k_\ell} + 6\frac{q_{1,2}}{k_q}+3\frac{q_3}{k_q}\,,\qquad\tilde\mu_\text{ss} = - 4\frac{u}{k_u}-\frac{t}{k_t} - \frac{b}{k_b} +4\frac{q_{1,2}}{k_q}+2\frac{q_3}{k_q}\,.
\end{split}
\end{align}
Here, $u$ is a proxy for the RH $u$, $d$, $c$ and $s$ quark asymmetries. The values used for the newly introduced quantities $D_\tau$, $\Gamma_Y^\tau$, $k_\tau$, $k_t$, $k_b$, and $\Gamma_\text{ss}$ (the strong sphaleron rate) are given in \cref{sec:app-inputs}.
The diffusion coefficient of the right handed quarks is identical to the one of the left handed quarks, $D_q$~\cite{Joyce:1994zn}.
Note that the net baryon number is now given by $B=\frac{1}{3}(4u+t+b+2q_{1,2}+q_3)$.
For our exemplary benchmark point presented in \cref{fig:solution}, the final $Y_B$ changes only by a few per cent when using the full transport equations.

%-----------------------------------------------------------------------------
\section{The Diffusion Coefficients}
\label{sec:app-diff-coeff}
%-----------------------------------------------------------------------------

The diffusion coefficient $D_a$ enters the transport equations, \cref{eq:transport-eq-dim-6,eq:transport-eq-dim-8}, through the divergence in \cref{eq:fokker-planck}. It measures the mobility of a species $a$ in the thermal bath and thus depends inversely on the scattering rate with the plasma. For the SM particles and $N_R$, we use diffusion coefficients found in the literature~\cite{Long:2017rdo,Joyce:1994zn,Cline:2000nw,Cline:2001rk}, i.e.,
\begin{align}
    D_N \approx \frac{16\pi^2}{T}\frac{1}{y_\nu^4} \, \qquad
    D_\tau \approx \frac{380}{T} \,, \qquad
    D_\ell \approx D_h \approx \frac{100}{T} \,, \qquad 
    D_q \approx \frac{6}{T} \,. 
\end{align}

\begin{figure}
	\centering
	\includegraphics{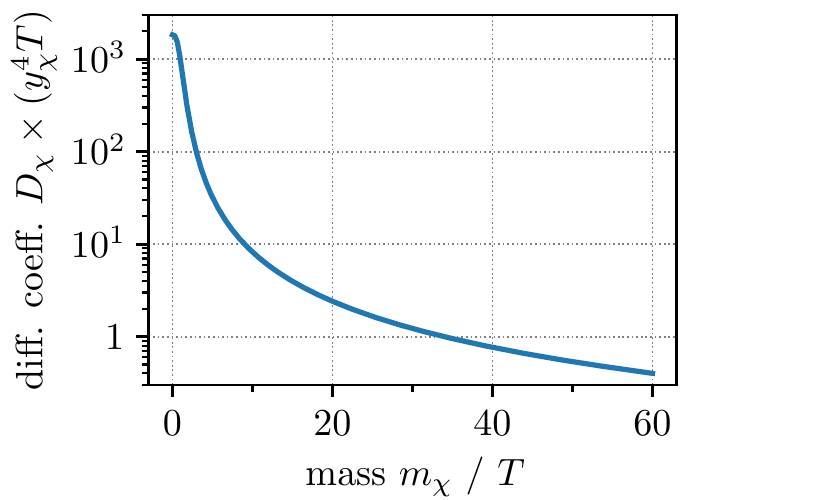}
	\caption{Mass dependent diffusion coefficient.\label{fig:diff-coeff}}
\end{figure}

In order to determine the diffusion coefficient for our dark matter particle $\chi$, we perform a calculation similar to the one given in ref.~\cite{Joyce:1994zn}.
In our case, however, $m_\chi(z)$ substantially increases during the dark phase transition, with $m_\chi \gg T$ in the broken phase.
We therefore have to consider a $z$-dependent diffusion coefficient, $D(z)$, which we evaluate numerically.
We start off with the spin-averaged squared matrix element for the scattering process $\chi\chi\to\chi\chi$, which consists of $t$- and $u$-channel diagrams mediated by a $\phi$:
\begin{multline}
    |\mathcal M|^2=y_\chi^4\ba{2\frac{(\frac{t}{4}-m_\chi^2(z))^2}{(t-m_\phi^2)^2}+2\frac{(\frac{u}{4}-m_\chi^2(z))^2}{(u-m_\phi^2)^2} \\
    +\frac{(\frac{s}{4}-m_\chi^2(z))^2-(\frac{t}{4}-m_\chi^2(z))^2-(\frac{u}{4}-m_\chi^2(z))^2}{(t-m_\phi^2)(u-m_\phi^2)}} \,.
\end{multline}
For the scalar mediator we assume a mass of $m_\phi = T_n$. The total interaction rate for a particle with 4-momentum $p$ and energy $E_p$ is~\cite{Joyce:1994zn}
\begin{align}
    \Gamma_\chi^\text{tot}(p) &= \frac{1}{2E_p}\int \upd \Pi_{k} \upd \Pi_{p'}\upd \Pi_{k'} (2\pi)^4 \delta^{(4)}(p+k-p'-k')|\mathcal M|^2\,(-f_\chi'(E_k))(1-\cos\theta) \,,
\end{align}
where $\upd\Pi_{k} \equiv \upd^3 k/(16\pi^3E_k)$ and $f_\chi'\equiv -T\frac{\partial f_\chi}{\partial E}$.
This rate can conveniently be evaluated in the centre-of-mass frame.
Finally, the diffusion coefficient is given by
\begin{align} \label{eq:diff-coeff}
    D_\chi(z)=\frac{12}{T^3}\int \frac{\upd^3 p}{(2\pi)^3} \frac{-f_\chi'(E_p)}{\Gamma_\chi^\text{tot}} \ba{\frac{p_z}{E_p}}^{\!2}\propto\frac{1}{y_\chi^4}\,.
\end{align}
\Cref{fig:diff-coeff} shows the diffusion coefficient as a function of mass.

%-----------------------------------------------------------------------------
\section{Input Parameters}
\label{sec:app-inputs}
%-----------------------------------------------------------------------------

\begin{table}
    \centering
    \begin{tabular}{p{3cm}r@{\;}lp{5.3cm}}
        \toprule
        Quantity & & Value & Reference\\
        \toprule
        \mbox{Diffusion coefficients}
            & $D_N\approx$ & $\frac{16\pi^2}{T}\frac{1}{y_\nu^4}$& \cite[p.~8]{Long:2017rdo} \\
            & $D_\tau\approx$ & $\frac{380}{T}$ & \cite[eq.~(130)]{Joyce:1994zn} \\
            & $D_\ell\approx D_h\approx$ & $\frac{100}{T}$ & ---\texttt{"}--- \\
            & $D_q\approx$ & $\frac{6}{T}$ & ---\texttt{"}--- \\
            & $D_\chi(z)\propto$ & $ \frac{1}{y_\chi^4}$ & \cref{eq:diff-coeff} \\
        \midrule
        \mbox{Yukawa \& relaxation rates}
            & $\Gamma_Y^N\approx$ & $\num{7.9e-3} \,y_\nu^2\,T$ & \cite[eq.~(137)]{Joyce:1994zn},\cite{Cirigliano:2006wh},\cite[eq.~(B.8)]{deVries:2017ncy} \\ 
            & $\Gamma_Y^\tau\approx$ & $\num{1.4e-6}\,T$ & ---\texttt{"}--- \\ 
            & $\Gamma_M^\chi(z)\approx$ & $\num{2.1e-1} \frac{m_\chi^2(z)}{T}$ & \cite[eq.~(15)]{Baier:1994wh},\cite[eq.~(2)]{Elmfors:1998hh},\newline\cite[eq.~(B.6)]{deVries:2017ncy},\cite[eq.~(A13)]{Fuchs:2020pun} \\
        \midrule
        \mbox{Sphaleron rates}
            & $\Gamma_\text{ws}\approx$ & $\num{5.2e-6}\,T$ & \cite[eq.~(A16)]{deVries:2017ncy}\\
            & $\Gamma_\text{ss}\approx$ & $\num{2.9e-3}\,T$ & \cite[(6.1)]{Moore:2010jd}\cite[eq.~(A16)]{deVries:2017ncy}\\
        \midrule
        \mbox{$k$-functions}      
            & $k_N\approx k_\tau\approx$ & 1 & \cite[eq.~(B.4)]{deVries:2017ncy},\cite[eq.~(A7)]{Fuchs:2020pun}\\
            & $k_\ell\approx$ & 2 & ---\texttt{"}---\\
            & $k_h\approx k_t\approx k_b\approx$ & 3 & ---\texttt{"}--- \\
            & $k_q\approx$ & 6 & ---\texttt{"}--- \\
            & $k_\chi(z\ll 0)\approx$ & 1 & ---\texttt{"}---  \\
            & $k_\chi(z\gg 0)\approx$ & 0 & ---\texttt{"}--- \\
        \midrule
        \mbox{Baryon asymmetry}
            & $Y_B^\text{obs}\approx$ & \num{8.65e-11} & \cite{Planck:2018vyg}\\
        \bottomrule
    \end{tabular}
    \caption{Summary of the physical quantities used in our Filtered Baryogenesis simulations, together with the appropriate references.}
    \label{tab:inputs}
\end{table}
In \cref{tab:inputs} we summarise the input parameters for our simulations, including the diffusion coefficients, the rates for various processes, and the $k$-factors.
For all expressions, we give the relevant references.

%-----------------------------------------------------------------------------
\section{Validity of the Diffusion Approximation} 
\label{sec:app-diff-approx}
%-----------------------------------------------------------------------------

As discussed in \cref{sec:results}, the diffusion ansatz is no longer justified if the mean free path of $\chi$ is not the shortest relevant length scale.
\Cref{tab:inputs} shows that the shortest length scale is associated with the inverse chiral relaxation rate $1/\Gamma_M^\chi$.  
This means that the change in the asymmetry due to chiral relaxation over one diffusion length is relatively large, which invalidates the diffusion ansatz and would require a full kinematic description~\cite{Joyce:1994bi}.

\begin{figure}
	\centering
	\hspace{0.35cm}	Dimension-6 $N_R$ portal \hspace{3cm} Dimension-8 Weinberg portal \\[0.2cm]
    \includegraphics{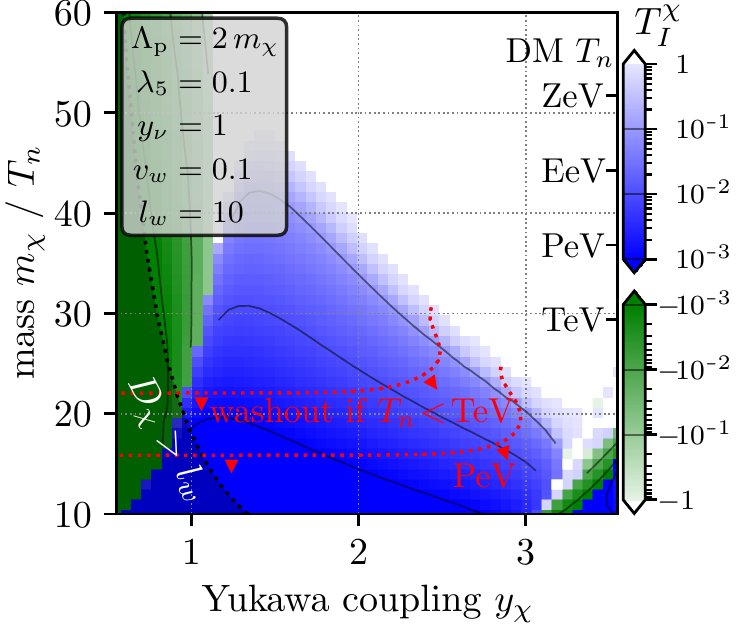} \hfill
    \includegraphics{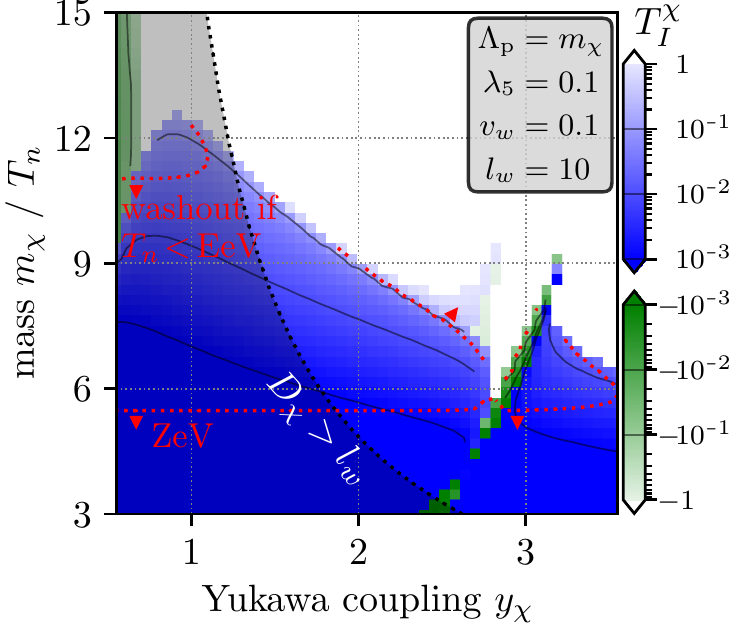}
	\caption{Parameter space of the Filtered Baryogenesis scenario with the dimension-6 $N_R$ portal (left) and with the dimension-8 Weinberg portal (right). Compared to \cref{fig:paramspace}, we use here a modified version of the chiral relaxation rate $\Gamma_M^\chi$ (see \cref{eq:GammaM-tilde}) to account for the fact that the chiral asymmetry will be generated at momenta much larger than the temperature.}
	\label{fig:paramspace-modgamma}
\end{figure}

$\Gamma_M^\chi$ is proportional to a momentum integral over a combination of equilibrium distribution functions (see, e.g., eq.~(B6) of ref.~\cite{deVries:2017ncy}).
This is based on the assumption that particles thermalise rapidly, implying that any chiral asymmetry that is generated quickly gets evenly distributed across momentum modes.
In our scenario, however, we expect the $\chi$ asymmetry to arise predominantly at momenta close to the reflection threshold, $p_z \sim m_\chi \gg T_n$.
If $D_\chi \gtrsim 1/\Gamma_M^\chi$, $\chi$ does not equilibrate efficiently, but rather stays concentrated in these momentum modes.
To account for this, we define a modified chiral relaxation rate as
\begin{align}
    \tilde \Gamma_M^\chi(z) \equiv  \Gamma_M^\chi(z) \frac{\mathcal I\Big(\sqrt{m^2_\chi-m_\chi^2(z)}\Big)}{\mathcal I(T_n)}\,,
    \label{eq:GammaM-tilde}
\end{align}
where $\mathcal I(k)$ is the integrand in the definition of $\Gamma_M^\chi(z)$ in ref.~\cite{deVries:2017ncy}, without the $k^2$ Jacobian, and divided by a Fermi-Dirac distribution (in order to obtain an expression that is not exponentially dependent on the integral momentum $k$). 
The modified relaxation rate is in general smaller than the original one.
In \cref{fig:paramspace-modgamma} we present the same slices of the parameter space as were shown in \cref{fig:paramspace}, but using $\tilde \Gamma_M^\chi$ instead of $\Gamma_M^\chi$.
The modification deforms the regions of successful baryogenesis.
A full kinematic treatment would likely yield results somewhere in between the two approaches.
We conclude that despite the limitations of the diffusion approach, Filtered Baryogenesis remains a viable mechanism in the considered parameter space.

%%%%%%%%%%%%%%%%%%%%%%%%%%%%%%%%%%%
\bibliographystyle{JHEP}
\bibliography{refs}

\providecommand{\href}[2]{#2}\begingroup\raggedright\begin{thebibliography}{10}

\bibitem{Baker:2019ndr}
M.J.~Baker, J.~Kopp and A.J.~Long, \emph{{Filtered Dark Matter at a First Order
  Phase Transition}},
  \href{https://doi.org/10.1103/PhysRevLett.125.151102}{\emph{Phys. Rev. Lett.}
  {\bfseries 125} (2020) 151102}
  [\href{https://arxiv.org/abs/1912.02830}{{\ttfamily 1912.02830}}].

\bibitem{Chway:2019kft}
D.~Chway, T.H.~Jung and C.S.~Shin, \emph{{Dark matter filtering-out effect
  during a first-order phase transition}},
  \href{https://doi.org/10.1103/PhysRevD.101.095019}{\emph{Phys. Rev. D}
  {\bfseries 101} (2020) 095019}
  [\href{https://arxiv.org/abs/1912.04238}{{\ttfamily 1912.04238}}].

\bibitem{Battaglieri:2017aum}
M.~Battaglieri et~al., \emph{{US Cosmic Visions: New Ideas in Dark Matter 2017:
  Community Report}},  in \emph{{U.S. Cosmic Visions: New Ideas in Dark
  Matter}}, 7, 2017 [\href{https://arxiv.org/abs/1707.04591}{{\ttfamily
  1707.04591}}].

\bibitem{Kuzmin:1985mm}
V.A.~Kuzmin, V.A.~Rubakov and M.E.~Shaposhnikov, \emph{{On the Anomalous
  Electroweak Baryon Number Nonconservation in the Early Universe}},
  \href{https://doi.org/10.1016/0370-2693(85)91028-7}{\emph{Phys. Lett.}
  {\bfseries B155} (1985) 36}.

\bibitem{Shaposhnikov:1986jp}
M.E.~Shaposhnikov, \emph{{Possible Appearance of the Baryon Asymmetry of the
  Universe in an Electroweak Theory}}, {\emph{JETP Lett.} {\bfseries 44} (1986)
  465} [Pisma Zh. Eksp. Teor. Fiz.44,364(1986)].

\bibitem{Shaposhnikov:1987tw}
M.E.~Shaposhnikov, \emph{{Baryon Asymmetry of the Universe in Standard
  Electroweak Theory}},
  \href{https://doi.org/10.1016/0550-3213(87)90127-1}{\emph{Nucl. Phys.}
  {\bfseries B287} (1987) 757}.

\bibitem{Carena:1996wj}
M.~Carena, M.~Quiros and C.E.M.~Wagner, \emph{{Opening the window for
  electroweak baryogenesis}},
  \href{https://doi.org/10.1016/0370-2693(96)00475-3}{\emph{Phys. Lett.}
  {\bfseries B380} (1996) 81}
  [\href{https://arxiv.org/abs/hep-ph/9603420}{{\ttfamily hep-ph/9603420}}].

\bibitem{Huber:2006wf}
S.J.~Huber, T.~Konstandin, T.~Prokopec and M.G.~Schmidt, \emph{{Electroweak
  Phase Transition and Baryogenesis in the nMSSM}},
  \href{https://doi.org/10.1016/j.nuclphysb.2006.09.003}{\emph{Nucl. Phys.}
  {\bfseries B757} (2006) 172}
  [\href{https://arxiv.org/abs/hep-ph/0606298}{{\ttfamily hep-ph/0606298}}].

\bibitem{Vaskonen:2016yiu}
V.~Vaskonen, \emph{{Electroweak baryogenesis and gravitational waves from a
  real scalar singlet}},  [\href{https://arxiv.org/abs/1611.02073}{{\ttfamily
  1611.02073}}].

\bibitem{Garbrecht:2018mrp}
B.~Garbrecht, \emph{{Why is there more matter than antimatter? Calculational
  methods for leptogenesis and electroweak baryogenesis}},
  [\href{https://arxiv.org/abs/1812.02651}{{\ttfamily 1812.02651}}].

\bibitem{deVries:2018tgs}
J.~De~Vries, M.~Postma and J.~van~de Vis, \emph{{The role of leptons in
  electroweak baryogenesis}},
  \href{https://doi.org/10.1007/JHEP04(2019)024}{\emph{JHEP} {\bfseries 04}
  (2019) 024} [\href{https://arxiv.org/abs/1811.11104}{{\ttfamily
  1811.11104}}].

\bibitem{Cline:2020jre}
J.M.~Cline and K.~Kainulainen, \emph{{Electroweak baryogenesis at high wall
  velocities}},  [\href{https://arxiv.org/abs/2001.00568}{{\ttfamily
  2001.00568}}].

\bibitem{Laurent:2020gpg}
B.~Laurent and J.M.~Cline, \emph{{Fluid equations for fast-moving electroweak
  bubble walls}},
  \href{https://doi.org/10.1103/PhysRevD.102.063516}{\emph{Phys. Rev. D}
  {\bfseries 102} (2020) 063516}
  [\href{https://arxiv.org/abs/2007.10935}{{\ttfamily 2007.10935}}].

\bibitem{Fuchs:2020pun}
E.~Fuchs, M.~Losada, Y.~Nir and Y.~Viernik, \emph{{Analytic Techniques for
  Solving the Transport Equations in Electroweak Baryogenesis}},
  [\href{https://arxiv.org/abs/2007.06940}{{\ttfamily 2007.06940}}].

\bibitem{Cohen:1993nk}
A.G.~Cohen, D.B.~Kaplan and A.E.~Nelson, \emph{{Progress in electroweak
  baryogenesis}},
  \href{https://doi.org/10.1146/annurev.ns.43.120193.000331}{\emph{Ann. Rev.
  Nucl. Part. Sci.} {\bfseries 43} (1993) 27}
  [\href{https://arxiv.org/abs/hep-ph/9302210}{{\ttfamily hep-ph/9302210}}].

\bibitem{Cline:2006ts}
J.M.~Cline, \emph{{Baryogenesis}},  in \emph{{Les Houches Summer School -
  Session 86: Particle Physics and Cosmology: The Fabric of Spacetime}}, 9,
  2006 [\href{https://arxiv.org/abs/hep-ph/0609145}{{\ttfamily
  hep-ph/0609145}}].

\bibitem{Morrissey:2012db}
D.E.~Morrissey and M.J.~Ramsey-Musolf, \emph{{Electroweak baryogenesis}},
  \href{https://doi.org/10.1088/1367-2630/14/12/125003}{\emph{New J. Phys.}
  {\bfseries 14} (2012) 125003}
  [\href{https://arxiv.org/abs/1206.2942}{{\ttfamily 1206.2942}}].

\bibitem{Arakawa:2021wgz}
J.~Arakawa, A.~Rajaraman and T.M.P.~Tait, \emph{{Annihilogenesis}},
  [\href{https://arxiv.org/abs/2109.13941}{{\ttfamily 2109.13941}}].

\bibitem{Azatov:2021irb}
A.~Azatov, M.~Vanvlasselaer and W.~Yin, \emph{{Baryogenesis via relativistic
  bubble walls}}, \href{https://doi.org/10.1007/JHEP10(2021)043}{\emph{JHEP}
  {\bfseries 10} (2021) 043}
  [\href{https://arxiv.org/abs/2106.14913}{{\ttfamily 2106.14913}}].

\bibitem{Dutta:2006pt}
B.~Dutta and J.~Kumar, \emph{{Hidden sector baryogenesis}},
  \href{https://doi.org/10.1016/j.physletb.2006.09.069}{\emph{Phys. Lett. B}
  {\bfseries 643} (2006) 284}
  [\href{https://arxiv.org/abs/hep-th/0608188}{{\ttfamily hep-th/0608188}}].

\bibitem{Shelton:2010ta}
J.~Shelton and K.M.~Zurek, \emph{{Darkogenesis: A baryon asymmetry from the
  dark matter sector}},
  \href{https://doi.org/10.1103/PhysRevD.82.123512}{\emph{Phys. Rev. D}
  {\bfseries 82} (2010) 123512}
  [\href{https://arxiv.org/abs/1008.1997}{{\ttfamily 1008.1997}}].

\bibitem{Hall:2019ank}
E.~Hall, T.~Konstandin, R.~McGehee, H.~Murayama and G.~Servant,
  \emph{{Baryogenesis From a Dark First-Order Phase Transition}},
  \href{https://doi.org/10.1007/JHEP04(2020)042}{\emph{JHEP} {\bfseries 04}
  (2020) 042} [\href{https://arxiv.org/abs/1910.08068}{{\ttfamily
  1910.08068}}].

\bibitem{Cline:2017qpe}
J.M.~Cline, K.~Kainulainen and D.~Tucker-Smith, \emph{{Electroweak baryogenesis
  from a dark sector}},  [\href{https://arxiv.org/abs/1702.08909}{{\ttfamily
  1702.08909}}].

\bibitem{Choi:1993cv}
J.~Choi and R.R.~Volkas, \emph{{Real Higgs singlet and the electroweak phase
  transition in the Standard Model}},
  \href{https://doi.org/10.1016/0370-2693(93)91013-D}{\emph{Phys. Lett. B}
  {\bfseries 317} (1993) 385}
  [\href{https://arxiv.org/abs/hep-ph/9308234}{{\ttfamily hep-ph/9308234}}].

\bibitem{Espinosa:2011ax}
J.R.~Espinosa, T.~Konstandin and F.~Riva, \emph{{Strong Electroweak Phase
  Transitions in the Standard Model with a Singlet}},
  \href{https://doi.org/10.1016/j.nuclphysb.2011.09.010}{\emph{Nucl. Phys. B}
  {\bfseries 854} (2012) 592}
  [\href{https://arxiv.org/abs/1107.5441}{{\ttfamily 1107.5441}}].

\bibitem{Cline:2012hg}
J.M.~Cline and K.~Kainulainen, \emph{{Electroweak baryogenesis and dark matter
  from a singlet Higgs}},
  \href{https://doi.org/10.1088/1475-7516/2013/01/012}{\emph{JCAP} {\bfseries
  1301} (2013) 012} [\href{https://arxiv.org/abs/1210.4196}{{\ttfamily
  1210.4196}}].

\bibitem{Nussinov:1985xr}
S.~Nussinov, \emph{{TECHNOCOSMOLOGY: COULD A TECHNIBARYON EXCESS PROVIDE A
  'NATURAL' MISSING MASS CANDIDATE?}},
  \href{https://doi.org/10.1016/0370-2693(85)90689-6}{\emph{Phys. Lett. B}
  {\bfseries 165} (1985) 55}.

\bibitem{Barr:1990ca}
S.M.~Barr, R.S.~Chivukula and E.~Farhi, \emph{{Electroweak Fermion Number
  Violation and the Production of Stable Particles in the Early Universe}},
  \href{https://doi.org/10.1016/0370-2693(90)91661-T}{\emph{Phys. Lett. B}
  {\bfseries 241} (1990) 387}.

\bibitem{Barr:1991qn}
S.M.~Barr, \emph{{Baryogenesis, sphalerons and the cogeneration of dark
  matter}}, \href{https://doi.org/10.1103/PhysRevD.44.3062}{\emph{Phys. Rev. D}
  {\bfseries 44} (1991) 3062}.

\bibitem{Kaplan:2009ag}
D.E.~Kaplan, M.A.~Luty and K.M.~Zurek, \emph{{Asymmetric Dark Matter}},
  \href{https://doi.org/10.1103/PhysRevD.79.115016}{\emph{Phys. Rev. D}
  {\bfseries 79} (2009) 115016}
  [\href{https://arxiv.org/abs/0901.4117}{{\ttfamily 0901.4117}}].

\bibitem{An:2009vq}
H.~An, S.-L.~Chen, R.N.~Mohapatra and Y.~Zhang, \emph{{Leptogenesis as a Common
  Origin for Matter and Dark Matter}},
  \href{https://doi.org/10.1007/JHEP03(2010)124}{\emph{JHEP} {\bfseries 1003}
  (2010) 124} [\href{https://arxiv.org/abs/0911.4463}{{\ttfamily 0911.4463}}],
  16 pages, 1 figure/ references added.

\bibitem{Haba:2010bm}
N.~Haba and S.~Matsumoto, \emph{{Baryogenesis from Dark Sector}},
  \href{https://doi.org/10.1143/PTP.125.1311}{\emph{Prog.Theor.Phys.}
  {\bfseries 125} (2011) 1311}
  [\href{https://arxiv.org/abs/1008.2487}{{\ttfamily 1008.2487}}].

\bibitem{Davoudiasl:2010am}
H.~Davoudiasl, D.E.~Morrissey, K.~Sigurdson and S.~Tulin, \emph{{Hylogenesis: A
  Unified Origin for Baryonic Visible Matter and Antibaryonic Dark Matter}},
  \href{https://doi.org/10.1103/PhysRevLett.105.211304}{\emph{Phys.Rev.Lett.}
  {\bfseries 105} (2010) 211304}
  [\href{https://arxiv.org/abs/1008.2399}{{\ttfamily 1008.2399}}], Footnote 1
  restored after omission in v2/ no other corrections.

\bibitem{Buckley:2010ui}
M.R.~Buckley and L.~Randall, \emph{{Xogenesis}},
  \href{https://doi.org/10.1007/JHEP09(2011)009}{\emph{JHEP} {\bfseries 1109}
  (2011) 009} [\href{https://arxiv.org/abs/1009.0270}{{\ttfamily 1009.0270}}],
  11 pages, 4 figures.

\bibitem{Gu:2010ft}
P.-H.~Gu, M.~Lindner, U.~Sarkar and X.~Zhang, \emph{{WIMP Dark Matter and
  Baryogenesis}},
  \href{https://doi.org/10.1103/PhysRevD.83.055008}{\emph{Phys.Rev.} {\bfseries
  D83} (2011) 055008} [\href{https://arxiv.org/abs/1009.2690}{{\ttfamily
  1009.2690}}].

\bibitem{Blennow:2010qp}
M.~Blennow, B.~Dasgupta, E.~Fernandez-Martinez and N.~Rius, \emph{{Aidnogenesis
  via Leptogenesis and Dark Sphalerons}},
  \href{https://doi.org/10.1007/JHEP03(2011)014}{\emph{JHEP} {\bfseries 03}
  (2011) 014} [\href{https://arxiv.org/abs/1009.3159}{{\ttfamily 1009.3159}}].

\bibitem{Allahverdi:2010rh}
R.~Allahverdi, B.~Dutta and K.~Sinha, \emph{{Cladogenesis: Baryon-Dark Matter
  Coincidence from Branchings in Moduli Decay}},
  \href{https://doi.org/10.1103/PhysRevD.83.083502}{\emph{Phys.Rev.} {\bfseries
  D83} (2011) 083502} [\href{https://arxiv.org/abs/1011.1286}{{\ttfamily
  1011.1286}}].

\bibitem{Dutta:2010va}
B.~Dutta and J.~Kumar, \emph{{Asymmetric Dark Matter from Hidden Sector
  Baryogenesis}},
  \href{https://doi.org/10.1016/j.physletb.2011.04.036}{\emph{Phys. Lett. B}
  {\bfseries 699} (2011) 364}
  [\href{https://arxiv.org/abs/1012.1341}{{\ttfamily 1012.1341}}].

\bibitem{Falkowski:2011xh}
A.~Falkowski, J.T.~Ruderman and T.~Volansky, \emph{{Asymmetric Dark Matter from
  Leptogenesis}}, \href{https://doi.org/10.1007/JHEP05(2011)106}{\emph{JHEP}
  {\bfseries 1105} (2011) 106}
  [\href{https://arxiv.org/abs/1101.4936}{{\ttfamily 1101.4936}}].

\bibitem{Graesser:2011wi}
M.L.~Graesser, I.M.~Shoemaker and L.~Vecchi, \emph{{Asymmetric WIMP dark
  matter}}, \href{https://doi.org/10.1007/JHEP10(2011)110}{\emph{JHEP}
  {\bfseries 1110} (2011) 110}
  [\href{https://arxiv.org/abs/1103.2771}{{\ttfamily 1103.2771}}].

\bibitem{Buckley:2011kk}
M.R.~Buckley, \emph{{Asymmetric Dark Matter and Effective Operators}},
  \href{https://doi.org/10.1103/PhysRevD.84.043510}{\emph{Phys.Rev.} {\bfseries
  D84} (2011) 043510} [\href{https://arxiv.org/abs/1104.1429}{{\ttfamily
  1104.1429}}].

\bibitem{Bell:2011tn}
N.F.~Bell, K.~Petraki, I.M.~Shoemaker and R.R.~Volkas, \emph{{Pangenesis in a
  Baryon-Symmetric Universe: Dark and Visible Matter via the Affleck-Dine
  Mechanism}},
  \href{https://doi.org/10.1103/PhysRevD.84.123505}{\emph{Phys.Rev.} {\bfseries
  D84} (2011) 123505} [\href{https://arxiv.org/abs/1105.3730}{{\ttfamily
  1105.3730}}].

\bibitem{Cheung:2011if}
C.~Cheung and K.M.~Zurek, \emph{{Affleck-Dine Cogenesis}},
  \href{https://doi.org/10.1103/PhysRevD.84.035007}{\emph{Phys.Rev.} {\bfseries
  D84} (2011) 035007} [\href{https://arxiv.org/abs/1105.4612}{{\ttfamily
  1105.4612}}], 9 pages, no figures.

\bibitem{MarchRussell:2011fi}
J.~March-Russell and M.~McCullough, \emph{{Asymmetric Dark Matter via
  Spontaneous Co-Genesis}},
  \href{https://doi.org/10.1088/1475-7516/2012/03/019}{\emph{JCAP} {\bfseries
  1203} (2012) 019} [\href{https://arxiv.org/abs/1106.4319}{{\ttfamily
  1106.4319}}].

\bibitem{Cui:2011qe}
Y.~Cui, L.~Randall and B.~Shuve, \emph{{Emergent Dark Matter, Baryon, and
  Lepton Numbers}}, \href{https://doi.org/10.1007/JHEP08(2011)073}{\emph{JHEP}
  {\bfseries 1108} (2011) 073}
  [\href{https://arxiv.org/abs/1106.4834}{{\ttfamily 1106.4834}}].

\bibitem{Cui:2011ab}
Y.~Cui, L.~Randall and B.~Shuve, \emph{{A WIMPy Baryogenesis Miracle}},
  \href{https://doi.org/10.1007/JHEP04(2012)075}{\emph{JHEP} {\bfseries 04}
  (2012) 075} [\href{https://arxiv.org/abs/1112.2704}{{\ttfamily 1112.2704}}].

\bibitem{Davoudiasl:2012uw}
H.~Davoudiasl and R.N.~Mohapatra, \emph{{On Relating the Genesis of Cosmic
  Baryons and Dark Matter}},
  [\href{https://arxiv.org/abs/1203.1247}{{\ttfamily 1203.1247}}], Invited
  review for the New Journal of Physics focus issue on 'Origin of Matter'. 17
  pages, 2 figures.

\bibitem{Unwin:2012rp}
J.~Unwin, \emph{{Exodus: Hidden origin of dark matter and baryons}},
  [\href{https://arxiv.org/abs/1212.1425}{{\ttfamily 1212.1425}}].

\bibitem{Cui:2012jh}
Y.~Cui and R.~Sundrum, \emph{{Baryogenesis for WIMPs}},
  [\href{https://arxiv.org/abs/1212.2973}{{\ttfamily 1212.2973}}].

\bibitem{Barr:2013tea}
S.M.~Barr and H.-Y.~Chen, \emph{{Cogeneration of Dark Matter and Baryons by
  Non-Standard-Model Sphalerons in Unified Models}},
  \href{https://doi.org/10.1007/JHEP10(2013)129}{\emph{JHEP} {\bfseries 10}
  (2013) 129} [\href{https://arxiv.org/abs/1309.0020}{{\ttfamily 1309.0020}}].

\bibitem{Servant:2013uwa}
G.~Servant and S.~Tulin, \emph{{Higgsogenesis}},
  [\href{https://arxiv.org/abs/1304.3464}{{\ttfamily 1304.3464}}].

\bibitem{Feng:2013wn}
W.-Z.~Feng, A.~Mazumdar and P.~Nath, \emph{{Baryogenesis from dark matter in an
  inflationary universe}},  [\href{https://arxiv.org/abs/1302.0012}{{\ttfamily
  1302.0012}}].

\bibitem{Hall:2019rld}
E.~Hall, T.~Konstandin, R.~McGehee and H.~Murayama, \emph{{Asymmetric Matters
  from a Dark First-Order Phase Transition}},
  [\href{https://arxiv.org/abs/1911.12342}{{\ttfamily 1911.12342}}].

\bibitem{Hambye:2018qjv}
T.~Hambye, A.~Strumia and D.~Teresi, \emph{{Super-cool Dark Matter}},
  [\href{https://arxiv.org/abs/1805.01473}{{\ttfamily 1805.01473}}].

\bibitem{DelleRose:2019pgi}
L.~Delle~Rose, G.~Panico, M.~Redi and A.~Tesi, \emph{{Gravitational Waves from
  Supercool Axions}},
  \href{https://doi.org/10.1007/JHEP04(2020)025}{\emph{JHEP} {\bfseries 04}
  (2020) 025} [\href{https://arxiv.org/abs/1912.06139}{{\ttfamily
  1912.06139}}].

\bibitem{Akhmedov:1999uz}
E.K.~Akhmedov, \emph{{Neutrino physics}},  in \emph{{ICTP Summer School in
  Particle Physics}}, 6, 1999
  [\href{https://arxiv.org/abs/hep-ph/0001264}{{\ttfamily hep-ph/0001264}}].

\bibitem{Mohapatra:1986ks}
R.N.~Mohapatra and J.W.F.~Valle, \emph{{Solar Neutrino Oscillations From
  Superstrings}},
  \href{https://doi.org/10.1016/0370-2693(86)90011-0}{\emph{Phys. Lett. B}
  {\bfseries 177} (1986) 47}.

\bibitem{CentellesChulia:2020dfh}
S.~Centelles~Chuli\'a, R.~Srivastava and A.~Vicente, \emph{{The inverse seesaw
  family: Dirac and Majorana}},
  \href{https://doi.org/10.1007/JHEP03(2021)248}{\emph{JHEP} {\bfseries 03}
  (2021) 248} [\href{https://arxiv.org/abs/2011.06609}{{\ttfamily
  2011.06609}}].

\bibitem{Baker:2021nyl}
M.J.~Baker, M.~Breitbach, J.~Kopp and L.~Mittnacht, \emph{{Primordial Black
  Holes from First-Order Cosmological Phase Transitions}},
  [\href{https://arxiv.org/abs/2105.07481}{{\ttfamily 2105.07481}}].

\bibitem{Baker:2021sno}
M.J.~Baker, M.~Breitbach, J.~Kopp and L.~Mittnacht, \emph{{Detailed Calculation
  of Primordial Black Hole Formation During First-Order Cosmological Phase
  Transitions}},  [\href{https://arxiv.org/abs/2110.00005}{{\ttfamily
  2110.00005}}].

\bibitem{Matsumoto:2018acr}
S.~Matsumoto, Y.-L.S.~Tsai and P.-Y.~Tseng, \emph{{Light Fermionic WIMP Dark
  Matter with Light Scalar Mediator}},
  \href{https://doi.org/10.1007/JHEP07(2019)050}{\emph{JHEP} {\bfseries 07}
  (2019) 050} [\href{https://arxiv.org/abs/1811.03292}{{\ttfamily
  1811.03292}}].

\bibitem{Escudero:2016gzx}
M.~Escudero, A.~Berlin, D.~Hooper and M.-X.~Lin, \emph{{Toward (Finally!)
  Ruling Out Z and Higgs Mediated Dark Matter Models}},
  \href{https://doi.org/10.1088/1475-7516/2016/12/029}{\emph{JCAP} {\bfseries
  12} (2016) 029} [\href{https://arxiv.org/abs/1609.09079}{{\ttfamily
  1609.09079}}].

\bibitem{Djouadi:2011aa}
A.~Djouadi, O.~Lebedev, Y.~Mambrini and J.~Quevillon, \emph{{Implications of
  LHC searches for Higgs--portal dark matter}},
  \href{https://doi.org/10.1016/j.physletb.2012.01.062}{\emph{Phys. Lett. B}
  {\bfseries 709} (2012) 65} [\href{https://arxiv.org/abs/1112.3299}{{\ttfamily
  1112.3299}}].

\bibitem{deVries:2017ncy}
J.~de~Vries, M.~Postma, J.~van~de Vis and G.~White, \emph{{Electroweak
  Baryogenesis and the Standard Model Effective Field Theory}},
  \href{https://doi.org/10.1007/JHEP01(2018)089}{\emph{JHEP} {\bfseries 01}
  (2018) 089} [\href{https://arxiv.org/abs/1710.04061}{{\ttfamily
  1710.04061}}].

\bibitem{DeVries:2018aul}
J.~De~Vries, M.~Postma and J.~van~de Vis, \emph{{The role of leptons in
  electroweak baryogenesis}},
  \href{https://doi.org/10.1007/JHEP04(2019)024}{\emph{JHEP} {\bfseries 04}
  (2019) 024} [\href{https://arxiv.org/abs/1811.11104}{{\ttfamily
  1811.11104}}].

\bibitem{Fuchs:2020uoc}
E.~Fuchs, M.~Losada, Y.~Nir and Y.~Viernik, \emph{{$CP$ violation from $\tau$,
  $t$ and $b$ dimension-6 Yukawa couplings - interplay of baryogenesis, EDM and
  Higgs physics}}, \href{https://doi.org/10.1007/JHEP05(2020)056}{\emph{JHEP}
  {\bfseries 05} (2020) 056}
  [\href{https://arxiv.org/abs/2003.00099}{{\ttfamily 2003.00099}}].

\bibitem{Lee:2004we}
C.~Lee, V.~Cirigliano and M.J.~Ramsey-Musolf, \emph{{Resonant relaxation in
  electroweak baryogenesis}},
  \href{https://doi.org/10.1103/PhysRevD.71.075010}{\emph{Phys. Rev. D}
  {\bfseries 71} (2005) 075010}
  [\href{https://arxiv.org/abs/hep-ph/0412354}{{\ttfamily hep-ph/0412354}}].

\bibitem{Cirigliano:2006wh}
V.~Cirigliano, M.J.~Ramsey-Musolf, S.~Tulin and C.~Lee, \emph{{Yukawa and
  tri-scalar processes in electroweak baryogenesis}},
  \href{https://doi.org/10.1103/PhysRevD.73.115009}{\emph{Phys. Rev. D}
  {\bfseries 73} (2006) 115009}
  [\href{https://arxiv.org/abs/hep-ph/0603058}{{\ttfamily hep-ph/0603058}}].

\bibitem{Bodeker:1999gx}
D.~Bodeker, G.D.~Moore and K.~Rummukainen, \emph{{Chern-Simons number diffusion
  and hard thermal loops on the lattice}},
  \href{https://doi.org/10.1103/PhysRevD.61.056003}{\emph{Phys. Rev. D}
  {\bfseries 61} (2000) 056003}
  [\href{https://arxiv.org/abs/hep-ph/9907545}{{\ttfamily hep-ph/9907545}}].

\bibitem{Cline:2000kb}
J.M.~Cline and K.~Kainulainen, \emph{{A New source for electroweak baryogenesis
  in the MSSM}}, \href{https://doi.org/10.1103/PhysRevLett.85.5519}{\emph{Phys.
  Rev. Lett.} {\bfseries 85} (2000) 5519}
  [\href{https://arxiv.org/abs/hep-ph/0002272}{{\ttfamily hep-ph/0002272}}].

\bibitem{Kainulainen:2021oqs}
K.~Kainulainen, \emph{{CP-violating transport theory for electroweak
  baryogenesis with thermal corrections}},
  \href{https://doi.org/10.1088/1475-7516/2021/11/042}{\emph{JCAP} {\bfseries
  11} (2021) 042} [\href{https://arxiv.org/abs/2108.08336}{{\ttfamily
  2108.08336}}].

\bibitem{Cline:1997vk}
J.M.~Cline, M.~Joyce and K.~Kainulainen, \emph{{Supersymmetric electroweak
  baryogenesis in the WKB approximation}},
  \href{https://doi.org/10.1016/S0370-2693(97)01361-0}{\emph{Phys. Lett. B}
  {\bfseries 417} (1998) 79}
  [\href{https://arxiv.org/abs/hep-ph/9708393}{{\ttfamily hep-ph/9708393}}],
  [Erratum: Phys.Lett.B 448, 321--321 (1999)].

\bibitem{Cline:2000nw}
J.M.~Cline, M.~Joyce and K.~Kainulainen, \emph{{Supersymmetric electroweak
  baryogenesis}},
  \href{https://doi.org/10.1088/1126-6708/2000/07/018}{\emph{JHEP} {\bfseries
  07} (2000) 018} [\href{https://arxiv.org/abs/hep-ph/0006119}{{\ttfamily
  hep-ph/0006119}}].

\bibitem{Cline:2001rk}
J.M.~Cline, M.~Joyce and K.~Kainulainen, \emph{{Erratum for 'Supersymmetric
  electroweak baryogenesis'}},
  [\href{https://arxiv.org/abs/hep-ph/0110031}{{\ttfamily hep-ph/0110031}}].

\bibitem{Kainulainen:2001cn}
K.~Kainulainen, T.~Prokopec, M.G.~Schmidt and S.~Weinstock, \emph{{First
  principle derivation of semiclassical force for electroweak baryogenesis}},
  \href{https://doi.org/10.1088/1126-6708/2001/06/031}{\emph{JHEP} {\bfseries
  06} (2001) 031} [\href{https://arxiv.org/abs/hep-ph/0105295}{{\ttfamily
  hep-ph/0105295}}].

\bibitem{Kainulainen:2002th}
K.~Kainulainen, T.~Prokopec, M.G.~Schmidt and S.~Weinstock,
  \emph{{Semiclassical force for electroweak baryogenesis: Three-dimensional
  derivation}}, \href{https://doi.org/10.1103/PhysRevD.66.043502}{\emph{Phys.
  Rev. D} {\bfseries 66} (2002) 043502}
  [\href{https://arxiv.org/abs/hep-ph/0202177}{{\ttfamily hep-ph/0202177}}].

\bibitem{Kainulainen:2021zbf}
K.~Kainulainen and O.~Koskivaara, \emph{{Non-equilibrium dynamics of a scalar
  field with quantum backreaction}},
  \href{https://doi.org/10.1007/JHEP12(2021)190}{\emph{JHEP} {\bfseries 12}
  (2021) 190} [\href{https://arxiv.org/abs/2105.09598}{{\ttfamily
  2105.09598}}].

\bibitem{Fromme:2006wx}
L.~Fromme and S.J.~Huber, \emph{{Top transport in electroweak baryogenesis}},
  \href{https://doi.org/10.1088/1126-6708/2007/03/049}{\emph{JHEP} {\bfseries
  03} (2007) 049} [\href{https://arxiv.org/abs/hep-ph/0604159}{{\ttfamily
  hep-ph/0604159}}].

\bibitem{Cline:2021dkf}
J.M.~Cline and B.~Laurent, \emph{{Electroweak baryogenesis from light fermion
  sources: A critical study}},
  \href{https://doi.org/10.1103/PhysRevD.104.083507}{\emph{Phys. Rev. D}
  {\bfseries 104} (2021) 083507}
  [\href{https://arxiv.org/abs/2108.04249}{{\ttfamily 2108.04249}}].

\bibitem{Planck:2018vyg}
{\scshape Planck} collaboration, \emph{{Planck 2018 results. VI. Cosmological
  parameters}},
  \href{https://doi.org/10.1051/0004-6361/201833910}{\emph{Astron. Astrophys.}
  {\bfseries 641} (2020) A6}
  [\href{https://arxiv.org/abs/1807.06209}{{\ttfamily 1807.06209}}], [Erratum:
  Astron.Astrophys. 652, C4 (2021)].

\bibitem{Gutlein:2010tq}
A.~Gutlein et~al., \emph{{Solar and atmospheric neutrinos: Background sources
  for the direct dark matter search}},
  \href{https://doi.org/10.1016/j.astropartphys.2010.06.002}{\emph{Astropart.
  Phys.} {\bfseries 34} (2010) 90}
  [\href{https://arxiv.org/abs/1003.5530}{{\ttfamily 1003.5530}}].

\bibitem{Kopp:2012dz}
J.~Kopp, \emph{{New Signals in Dark Matter Detectors}},
  \href{https://doi.org/10.1088/1742-6596/485/1/012032}{\emph{J. Phys. Conf.
  Ser.} {\bfseries 485} (2014) 012032}
  [\href{https://arxiv.org/abs/1210.2703}{{\ttfamily 1210.2703}}].

\bibitem{Billard:2013qya}
J.~Billard, L.~Strigari and E.~Figueroa-Feliciano, \emph{{Implication of
  neutrino backgrounds on the reach of next generation dark matter direct
  detection experiments}},
  \href{https://doi.org/10.1103/PhysRevD.89.023524}{\emph{Phys. Rev. D}
  {\bfseries 89} (2014) 023524}
  [\href{https://arxiv.org/abs/1307.5458}{{\ttfamily 1307.5458}}].

\bibitem{Griest:1989wd}
K.~Griest and M.~Kamionkowski, \emph{{Unitarity Limits on the Mass and Radius
  of Dark Matter Particles}},
  \href{https://doi.org/10.1103/PhysRevLett.64.615}{\emph{Phys. Rev. Lett.}
  {\bfseries 64} (1990) 615}.

\bibitem{Apreda:2001us}
R.~Apreda, M.~Maggiore, A.~Nicolis and A.~Riotto, \emph{{Gravitational waves
  from electroweak phase transitions}},
  \href{https://doi.org/10.1016/S0550-3213(02)00264-X}{\emph{Nucl. Phys. B}
  {\bfseries 631} (2002) 342}
  [\href{https://arxiv.org/abs/gr-qc/0107033}{{\ttfamily gr-qc/0107033}}].

\bibitem{Ellis:2016jkw}
J.~Ellis, \emph{{TikZ-Feynman: Feynman diagrams with TikZ}},
  \href{https://doi.org/10.1016/j.cpc.2016.08.019}{\emph{Comput. Phys. Commun.}
  {\bfseries 210} (2017) 103}
  [\href{https://arxiv.org/abs/1601.05437}{{\ttfamily 1601.05437}}].

\bibitem{Joyce:1994zn}
M.~Joyce, T.~Prokopec and N.~Turok, \emph{{Nonlocal electroweak baryogenesis.
  Part 1: Thin wall regime}},
  \href{https://doi.org/10.1103/PhysRevD.53.2930}{\emph{Phys. Rev. D}
  {\bfseries 53} (1996) 2930}
  [\href{https://arxiv.org/abs/hep-ph/9410281}{{\ttfamily hep-ph/9410281}}].

\bibitem{Long:2017rdo}
A.J.~Long, A.~Tesi and L.-T.~Wang, \emph{{Baryogenesis at a
  Lepton-Number-Breaking Phase Transition}},
  \href{https://doi.org/10.1007/JHEP10(2017)095}{\emph{JHEP} {\bfseries 10}
  (2017) 095} [\href{https://arxiv.org/abs/1703.04902}{{\ttfamily
  1703.04902}}].

\bibitem{Baier:1994wh}
R.~Baier and R.~Kobes, \emph{{On the damping rate of a fast fermion in hot
  QED}}, \href{https://doi.org/10.1103/PhysRevD.50.5944}{\emph{Phys. Rev. D}
  {\bfseries 50} (1994) 5944}
  [\href{https://arxiv.org/abs/hep-ph/9403335}{{\ttfamily hep-ph/9403335}}].

\bibitem{Elmfors:1998hh}
P.~Elmfors, K.~Enqvist, A.~Riotto and I.~Vilja, \emph{{Damping rates in the
  MSSM and electroweak baryogenesis}},
  \href{https://doi.org/10.1016/S0370-2693(99)00169-0}{\emph{Phys. Lett. B}
  {\bfseries 452} (1999) 279}
  [\href{https://arxiv.org/abs/hep-ph/9809529}{{\ttfamily hep-ph/9809529}}].

\bibitem{Moore:2010jd}
G.D.~Moore and M.~Tassler, \emph{{The Sphaleron Rate in SU(N) Gauge Theory}},
  \href{https://doi.org/10.1007/JHEP02(2011)105}{\emph{JHEP} {\bfseries 02}
  (2011) 105} [\href{https://arxiv.org/abs/1011.1167}{{\ttfamily 1011.1167}}].

\bibitem{Joyce:1994bi}
M.~Joyce, T.~Prokopec and N.~Turok, \emph{{Efficient electroweak baryogenesis
  from lepton transport}},
  \href{https://doi.org/10.1016/0370-2693(94)91377-3}{\emph{Phys. Lett. B}
  {\bfseries 338} (1994) 269}
  [\href{https://arxiv.org/abs/hep-ph/9401352}{{\ttfamily hep-ph/9401352}}].

\end{thebibliography}\endgroup
%%%%%%%%%%%%%%%%%%%%%%%%%%%%%%%%%%%

\end{document}